\documentclass{emulateapj}

\newcommand{\ha}{\hbox{H$\alpha$}}
\newcommand{\hi}{\hbox{H {\sc i}}}
\newcommand{\hii}{\hbox{H {\sc ii}}}
\newcommand{\oii}{\hbox{[O {\sc ii}]}}
\newcommand{\oiii}{\hbox{[O {\sc iii}]}}
\newcommand{\kms}{\hbox{$\rm {km}~\rm s^{-1}$}}

\newcommand{\sini}{\hbox{${\rm sin}~i$}}
\newcommand{\sigoned}{\hbox{$\sigma_{1d}$}}
\newcommand{\logsigoned}{\hbox{${\rm log}~\sigma_{1d}$}}
\newcommand{\sigtwod}{\hbox{$\sigma_{2d}$}}
\newcommand{\vrot}{\hbox{$V_{rot}$}}

\newcommand{\wone}{\hbox{$S_{1.0}$}}
\newcommand{\whalf}{\hbox{$S_{0.5}$}}
\newcommand{\etal}{et al.\,}
\newcommand{\mydeg}{\hbox{$^\circ$}}

\long\def\hidetext#1{\relax}

\shortauthors{Weiner et al.}
\shorttitle{Survey of Galaxy Kinematics. I. Rotation and Dispersion}

\slugcomment{Accepted by ApJ}

\begin{document}

\title{A Survey of Galaxy Kinematics to $z \sim 1$ in the TKRS/GOODS-N 
Field. \\ I. Rotation and Dispersion Properties
\altaffilmark{1}}

\author{Benjamin J. Weiner\altaffilmark{2}, 
Christopher N.A. Willmer\altaffilmark{3,4}, 
S.M. Faber\altaffilmark{5}, 
Jason Melbourne\altaffilmark{5},
Susan A. Kassin\altaffilmark{5},
Andrew C. Phillips\altaffilmark{5},
Justin Harker\altaffilmark{5},
A.J. Metevier\altaffilmark{6,7}, 
N.P. Vogt\altaffilmark{8},
D.C. Koo\altaffilmark{5}
}

\altaffiltext{1}{Based in part on observations taken at the W.M. Keck
Observatory, which is operated jointly by the University of California
and the California Institute of Technology}
\altaffiltext{2}{Department of Astronomy, University
of Maryland, College Park, MD 20742, {\tt bjw@astro.umd.edu}
Present address: Steward Observatory, University of Arizona, 
933 N. Cherry Av., Tucson, AZ 85721}
\altaffiltext{3}{Steward Observatory, University of Arizona, 
933 N. Cherry Av., Tucson, AZ 85721}
\altaffiltext{4}{On leave from Observatorio Nacional, Rio de Janeiro, Brasil}
\altaffiltext{5}{UCO/Lick Observatory, University of California, Santa Cruz, 
Santa Cruz, CA 95064}
\altaffiltext{6}{Center for Adaptive Optics, University of California, 
Santa Cruz, Santa Cruz, CA 95064}
\altaffiltext{7}{NSF Astronomy and Astrophysics Postdoctoral Fellow}
\altaffiltext{8}{Department of Astronomy, New Mexico State University,
P. O. Box 30001, Las Cruces, NM 88003}

\begin{abstract}

We present kinematic measurements of a large sample of galaxies from
the Team Keck Redshift Survey in the GOODS-N field.  We measure
line-of-sight velocity dispersions from linewidths of integrated emission 
for 1089 galaxies with median redshift 0.637, and spatially resolved
kinematics for a subsample of 380 galaxies.  This is the largest sample of
galaxies to $z \sim 1$ with kinematics to date, and allows us to
measure kinematic properties without morphological pre-selection.
Emission linewidths provide a dynamical measurement for the bulk of
the blue galaxy population.  To fit the spatially resolved kinematics,
we construct models with both line-of-sight rotation amplitude and velocity
dispersion as fit parameters.  Integrated linewidth correlates well
with a combination of the spatially-resolved velocity gradient and
dispersion, and is a robust measure of galaxy kinematics.  The spatial
extents of emission and continuum are similar and there is no evidence
that linewidths are affected by nuclear or clumpy emission.  The
measured rotation gradient is a strong function of slit position angle
alignment with galaxy major axis, but integrated linewidth is largely
independent of slit alignment.  Even in a subsample of galaxies with
well-aligned slits, there are galaxies whose kinematics are dominated
by dispersion ($V/\sigma<1$) rather than rotation.  These are probably
objects with disordered velocity fields, not dynamically hot stellar
systems.  About 35\% of the spatially resolved sample are dispersion
dominated; galaxies that are both dispersion dominated and bright
exist at high redshift but appear rare at low redshift.  This
kinematic morphology may yield a probe of the evolutionary state of
these objects.  Kinematic morphology is linked to photometric
morphology in HST/ACS images: dispersion dominated galaxies include a
higher fraction of irregulars and chain galaxies, while rotation
dominated galaxies are mostly disks and irregulars.  Only one-third of
chain/hyphen galaxies are dominated by rotation; high redshift
elongated objects cannot be assumed to be inclined disks.  In a
companion paper, we use the linewidths and rotation to measure
evolution in the Tully-Fisher relation.

\end{abstract}


\section{Introduction}

Galaxy redshift surveys to $z\sim 1$ allow the measurement
of evolution in properties of the galaxy population at
lookback times of half the present age of the Universe.
To date, a number
of surveys have studied galaxy properties such as luminosities,
colors, radii, morphology and environment.  The internal
kinematics of galaxies are equally interesting, as a 
physical probe of individual objects and to allow the
measurement of scaling relations of the galaxy population.  
However, measuring kinematics requires
moderate to high-resolution spectroscopy, which has
until recently been difficult to obtain for large numbers
of faint galaxies.

This paper measures kinematics of galaxies from emission
lines, using Keck/DEIMOS spectra from 
the Team Keck Redshift Survey (TKRS) in the GOODS-N 
(Great Observatories Origins Deep Survey) field 
(Wirth \etal\ 2004; Giavalisco \etal\ 2004).  The TKRS 
provides spectra of 1437 galaxies drawn from a magnitude 
limited sample to $R_{AB} < 24.4$, at resolution $R \sim 2100$.  
We select 1191 galaxies with good emission 
line detections and measure 1089 emission
line velocity dispersions from integrated emission,
with median $<z> = 0.637$.  
A subsample of 464 are selected
for modeling spatially resolved rotation and dispersion profiles, 
of which 380 yield good-quality measurements.  We use these spatially
resolved measures to probe the kinematic properties of high
redshift galaxies and to test the integrated linewidths.
In Paper II (Weiner \etal\ 2006), we use the linewidths and rotation
velocities to measure evolution in the Tully-Fisher relation.

A number of previous works have observed galaxy internal 
kinematics from $0.1<z<1.0$, for the purpose of measuring
Tully-Fisher relations.
The pioneering
studies of Vogt \etal\ (1996, 1997) modeled rotation curves for
17 galaxies of disky morphology with median $<z> = 0.47$
by combining Keck/LRIS slitlet spectra with structural information from 
HST photometry.  Several subsequent studies of rotation curves with similar
modeling procedures have contained 20-100 galaxies with 
median redshifts $\sim 0.4 - 0.5$ (e.g. Simard \& Pritchet 1998;
Vogt 2000; Ziegler \etal\ 2002; Milvang-Jensen et al.\ 2003; 
B{\"o}hm et al.\ 2004; Bamford \etal\ 2005,2006; 
Conselice \etal\ 2005; Nakamura et al.\ 2006; Metevier \etal\ 2006).
Generally these samples have been morphologically 
selected to be inclined disky objects, of which a majority
show measurable rotation.

Here we study a sample with emission line kinematics 
that is essentially selected only
on magnitude and emission line strength.  
We measure one-dimensional integrated velocity
dispersion (linewidth), and for a subsample, spatially resolved rotation 
profiles; we discuss the properties of these velocity measures in 
Sections \ref{sec-kinprops} and \ref{sec-spatialkin}.  We
show that the integrated linewidth is a fairly robust measure of the
characteristic velocity of a galaxy, expanding the scope of
the study beyond galaxies that are selected to be
orderly rotating disks.  Indeed, we find from the
spatially resolved rotation profiles that a significant 
number of high-redshift galaxies show kinematics that do not
appear to be orderly rotation.

We adopt a LCDM cosmology with $h=0.7$, $\Omega_M = 0.3$,
and $\Omega_\Lambda = 0.7$.  Magnitudes quoted in this paper are
in the AB system unless explicitly indicated as Vega.  
Section \ref{sec-observations} 
discusses the sample
and measurement methods, and section \ref{sec-complete}
presents the properties of the sample and completeness.  
Sections \ref{sec-kinprops} and \ref{sec-spatialkin}
study the simulated and empirical
properties of the kinematic measurements: integrated linewidths
and spatially resolved velocities.  
In Paper II we use the kinematics to measure
the evolution in the Tully-Fisher relation with redshift.

\section{Sample, Velocity Measurements, and Photometry}
\label{sec-observations}

\subsection{Properties of the spectroscopic sample}
\label{sec-sample}

The parent sample for our kinematic measurements is the 
Team Keck Redshift Survey (TKRS) in the GOODS-N field.
The selection and observations
are described at length by Wirth \etal (2004) and we 
present a few relevant details here.  Data and images of
TKRS galaxies may be retrieved from the TKRS website
at {\tt http://www2.keck.hawaii.edu/science/tksurvey/}.

The TKRS sample is magnitude selected to $R_{AB} = 24.4$.
No color or morphological selection was applied, but there
is a mild surface brightness selection (see Willmer \etal\ 2006),
which may eliminate a few faint high-$z$ reddened edge-on galaxies.
The sampling completeness
(ratio of galaxies with spectra to galaxies meeting
selection criteria) is $\sim 75\%$ and the redshift success rate
is $\sim 70\%$ ($>80\%$ for $R_{AB}<23$).  
Most galaxies for which spectra were taken but
redshifts could not be determined are faint, blue, and
probably at $z>1.5$, where \oii\ $\lambda$ 3727 disappears into
the night-sky OH line forest.

TKRS observations used the Keck II telescope and 
DEIMOS spectrograph with a 600 lines/mm grating, yielding
a sampling of 0.119\arcsec/pixel and 0.648 \AA/pixel.
Typical wavelength coverage was 4600--9800 \AA.
18 slitmasks were observed, each for 60 minutes total 
exposure time.  The slit width was 1.0\arcsec.  In the mask designs, 
slit position angles (PAs)
could be tilted to follow the major axis of elongated
objects, by up to 30 degrees away from the nominal
perpendicular-to-dispersion direction.  Galaxy PAs
were measured from ground-based $R$-band imaging before mask
design.  When slits were tilted, the slit width was adjusted to keep 
the width along the dispersion direction at 1.0\arcsec, keeping the
spectral resolution constant regardless of slit PA.

The TKRS data were reduced and redshifts measured with the DEIMOS 
pipeline software developed by the DEEP project, described in 
Davis \etal\ (2003) and Wirth \etal\ (2004); TKRS data are very similar
to DEEP2 data, save that DEEP2 uses a 1200 lines/mm grating.  
Candidate redshifts were measured automatically and verified by 
visual inspection by members of Team Keck.  

End data products include 2-d spectra for each slitlet and both
optimal and boxcar extracted 1-d spectra for each object.  We used the
2-d spectra to measure rotation curves and the 1-d spectra to measure
linewidths of integrated emission.  In this paper we always use the
boxcar extraction, never the optimal: the optimal extraction weights
different regions of the galaxy light profile unequally, which makes
the physical meaning of the optimally extracted spectrum somewhat
obscure.  The boxcar extraction window diameter is set by the reduction
software to 1.5 times the FWHM of the object measured in the 2-d
spectrum.  For nearly all galaxies, larger extraction windows do not
make a significant difference to the measured linewidths, because the
emission intensity is centrally peaked (see also Section
\ref{sec-blurdisk}).

\subsection{Emission line measurement in the 1-d spectra}
\label{sec-linefit}

Our goal in fitting linewidths to integrated emission is to
obtain a kinematic measurement for as many galaxies as possible,
without applying cuts to the sample.  Several previous studies
have used linewidths of integrated emission to measure kinematics
of small samples of galaxies at intermediate redshift 
(Rix \etal\ 1997; Mallen-Ornelas \etal\ 1999).

\subsubsection{Line fitting}

We developed an automated program to fit emission lines in all TKRS 
galaxies with secure redshifts (quality code 3 or 4 indicating
two spectral features, Wirth \etal\ 2004).  The {\sc LINEFIT}
program takes a galaxy redshift and list of common emission lines;
at the predicted location of each line, it
takes the wavelength, flux, and error data from a 40 \AA\ window
and fits a gaussian profile, using a Levenberg-Marquardt
non-linear least squares $\chi^2$ minimization (Press \etal\ 1992).

The fit has four free parameters: continuum level, line intensity,
velocity, and velocity dispersion.  For the \oii\ $\lambda\lambda$ 
3726.0, 3728.8 doublet, the program can fit a doublet with the 
wavelength ratio fixed and the intensity ratio fixed or free;
for the TKRS data, we fixed the doublet intensity ratio at 1.4 
(the mean ratio of red/blue components in high-S/N data measured
with the 1200 lines/mm grating).
The least squares fitter yields best-fit values
and error estimates for all parameters.
Fitting in the 40 \AA\ window does not provide an adequate measure of
the continuum in low-S/N spectra and so {\sc LINEFIT} also measures 
a robust continuum by taking the biweight of data in two 80 \AA\
windows on either side of the emission line.

\subsubsection{Instrumental resolution}

This sample relies heavily on the measurement of velocity
dispersion, which in turn can be strongly affected by the 
instrumental spectral resolution $\sigma_{inst}$.  Conventionally,
the restframe intrinsic line-of-sight velocity dispersion 
\sigoned\ of a line is given by 

\begin{equation}
\sigoned = \frac{c}{\lambda_{obs}} \sqrt{\sigma_{obs}^2 - \sigma_{inst}^2},
\label{eqn-siginst}
\end{equation}

\noindent 
where \sigoned\ is in \kms, $\lambda_{obs}$ is the observed
wavelength in \AA, and $\sigma_{obs}$ and $\sigma_{inst}$ are the
measured line dispersion and instrumental resolution, both in \AA.  
As a rule of thumb, measurements for $\sigoned < c\sigma_{inst}/\lambda_{obs}$
are not very reliable because small errors in the
observed width have a large effect on the inferred dispersion.
We use \sigoned\ to denote the restframe velocity dispersion derived
from the 1-dimensional extracted spectrum, which is integrated over
the extraction window.  Throughout these papers, observed quantities 
are given in \AA\ and restframe velocity quantities in \kms.  
This paper refers to several dispersion and velocity quantities,
in observed and restframe, and 1-d and 2-d spectra; these are
summarized in Table \ref{table-velquant}.

\begin{deluxetable}{lrr}

\tablecaption{
Dispersion and velocity-related quantities
\label{table-velquant}
}

\tablecolumns{3}
\tablewidth{0pt}

\tablehead{
  Quantity  & units  & description 
 }
 \startdata
$\sigma_{obs}$   &  \AA   &  Line dispersion measured in 1-d spectrum \\
$\sigma_{inst}$  &  \AA   &  Instrumental resolution \\
\sigoned         &  \kms  &  Intrinsic, instrumental-subtracted line-of-sight\\
                 &        &  restframe dispersion in 1-d spectrum \\
\vrot            &  \kms  &  Restframe line-of-sight rotation velocity \\
                 &        &  from modeling of 2-d spectrum \\
$\vrot/\sini$    &  \kms  &  Inclination-corrected rotation velocity \\
\sigtwod         &  \kms  &  Restframe line-of-sight intrinsic dispersion \\
                 &        &  from modeling of 2-d spectrum \\
$S_K$            &  \kms  &  Combined velocity, $\sqrt{K\vrot^2 + \sigtwod^2}$ \\
$V_c$            &  \kms  &  True circular rotation velocity \\

 \enddata

  \end{deluxetable}

With DEIMOS and 1.0\arcsec\ slits, the profile of night sky lines or
calibration arcs is somewhat flat-topped, less peaked than a 
gaussian.  This flat-topping is pronounced for night sky lines in 
the 600 lines/mm grating data of the TKRS, but is relatively small 
for DEIMOS spectra taken with the 1200 lines/mm grating, such as 
those in the DEEP2 survey (Davis \etal\ 2003).
Fitting gaussians to many line profiles of 
isolated night-sky lines in many different slits and masks yields
mean instrumental gaussian sigmas $\sigma_{inst,sky} = 1.4, 0.56$ \AA\
for TKRS and DEEP2 respectively.  

Flat-topping is less noticeable in the actual emission lines of 
observed galaxies in TKRS, for two reasons.  First, galaxies' velocity 
widths broaden the profile so that flat-topping is smeared out.
Second, TKRS galaxies are relatively small and their light 
distributions peak near the slit center, rather than uniformly
filling the slit as night sky lines do.  This slit underfilling
is potentially a significant problem for any study of velocity 
dispersions of small objects.  However, seeing and the relatively 
slow change of angular diameter with redshift beyond $z\sim 0.5$
mitigate the effect.  The effect is also smaller when the 
instrumental resolution is moderately high, as in TKRS and even 
more so for DEEP2.  

Modeling slit underfilling for the measured galaxy sizes
and comparing 
objects with both 600 and 1200 lines/mm observations shows that
flat-topping and its opposing effects nearly cancel out.
We matched observed linewidths for 70 galaxies on a DEEP2 slitmask
that was observed with both 600 and 1200 lines/mm gratings,
and found good agreement for 600 lines/mm $\sigma_{inst} = 1.4$ \AA.

The instrumental resolution as measured from night sky lines
varies by $\sim 5\%$ peak-peak over the DEIMOS field: lines from 
slitlets near the ends of the mask are broader than lines from 
slitlets near the center.  The effect is only visible statistically,
since its size is comparable to the scatter in individual 
sky line measurements and the scatter between different masks
(focus variations).  We have adopted a single value for
the instrumental resolution and neglected spatial and temporal
variations.  The variations' effect on the derived dispersions
is small: for a small galaxy with $\sigoned=50$ \kms, the peak
error induced is $\sim 2$ \kms, and the error declines rapidly
for larger \sigoned.

\subsubsection{Velocity dispersion sample}

There are 1437 galaxies with redshifts, TKRS spectra, and magnitudes.
We fit emission lines in each of these objects.  
For velocity dispersion purposes
we then reject all lines that do not have a 4 sigma
intensity detection.  The 4 sigma line catalog contains
2595 lines over 1191 galaxies.

Galaxies can have several lines, and the fit parameters are independent.  
To obtain one estimate of velocity dispersion for each object,
we take the weighted mean of the measurements of squared intrinsic velocity 
dispersion, $\sigoned^2$; using the square properly accounts for lines that 
are narrower than the nominal instrumental resolution, which have
$\sigoned^2<0$.
We use a weighted rms of the dispersions as the error estimate.
We exclude a tail of 85 galaxies that have both error$(\logsigoned)>0.25$
and error$(\sigoned)>30$ \kms\ to reject low-quality fits.

Some line fits have an observed width $\sigma_{obs}$
that is smaller than the
nominal instrumental width.  Although this is formally physically
impossible, it is expected in the presence of noisy data,
slit underfilling, and variations in the instrumental resolution.
After combining all lines, 196 of 1089 galaxies
are ``kinematically unresolved,'' with widths close to or less than 
instrumental, so that the formal velocity dispersion is undefined 
or has a large error in \logsigoned.
When we restrict to $M_B<-18$, the cutoff for our Tully-Fisher fits
in Paper II,
104 of 913 are kinematically unresolved in \logsigoned, under the
criteria ${\rm error}(\logsigoned)>0.25$, ${\rm error}(\sigoned)<30$,
and $\sigoned<25$ \kms.
Eliminating these galaxies from the sample preferentially
rejects low-velocity galaxies and leads to a bias,
so for plotting and fitting purposes we assign them a low value, 
$\logsigoned=1.4 \pm 0.2$.  The results of fitting do not
depend strongly on the exact value assigned.  Note that 
a few other galaxies have $\logsigoned<1.4$ yet low formal 
error on \logsigoned, usually because they have very strong and
well-measured emission lines.  In Section 3
of Paper II we outline a fitting method which 
treats the unresolved galaxies more robustly by fitting the
ensemble of observed width $\sigma_{obs}$ before the 
instrumental resolution is subtracted. 

Some galaxies with very bright emission lines can have formally
very small errors.  It is hubris to take these errors literally,
since we can hardly expect to measure velocity dispersions of 
distant galaxies to $<5\%$.  To be realistic and to prevent small
errors from dominating fits, we add 0.03 in quadrature to the error
on \logsigoned, to cut off the low tail of the error
distribution.  The results are not sensitive to the exact value used;
the median error for the sample of 913 brighter
than $M_B=-18$ is 0.084 in \logsigoned.  As discussed in Paper II,
the intrinsic scatter in \logsigoned\ dominates over the errors
on individual points.

\subsection{Rotation curve measurements}
\label{sec-rotcurve}

The DEIMOS slitlet spectra preserve spatial information and
it is common to see emission lines with a velocity gradient.
Spatially extended emission can be used to measure a rotation curve.
However, the spatial extent of emission is only a
few times the seeing, so ``beam smearing'' is large and the
effect of seeing must be modeled.  The rotation curve is
also affected by the inclination, slit width, and slit alignment 
with respect to the major axis of the galaxy.  We measured 
rotation curves for a subsample of objects in order to
study the relation between rotation velocity and integrated
linewidth.

\subsubsection{Previous high-redshift rotation curve modeling}

Previous efforts to measure rotation curves at intermediate to
high redshifts have approached these problems by modeling the
intrinsic emission intensity and velocity distribution,
subjecting them to seeing and slit effects, and fitting models
to the data to obtain a measure of the rotation velocity.
Vogt \etal\ (1996, 1997) pioneered work in this field, using
HST/WFPC2 imaging to model the galaxy light distribution and
parameters such as inclination and position angle, and fitting to 
Keck/LRIS slit spectra, varying the model rotation velocity.
Subsequent works in this field have followed similar modeling
programs, applying seeing and slit effects to a 2-d model of the
light and velocity fields (e.g. Simard \& Pritchet 1998; 
Ziegler \etal\ 2002; Milvang-Jensen et al.\ 2003; B{\"o}hm et al.\ 2004; 
Bamford \etal\ 2005, 2006; Conselice \etal\ 2005; Nakamura et al.\ 2006; 
Metevier \etal\ 2006).

These programs have generally focused on relatively small 
numbers of galaxies ($\sim 10-100$), in some cases selected
on morphology or spectra to be fairly normal rotating disks.  
They also require high-resolution imaging, preferably from HST, 
to obtain structural parameters.  
Most galaxies in the TKRS have deep HST/ACS imaging from the 
GOODS survey (Giavalisco \etal\ 2004), and full modeling of the
rotation curves is in progress.
However, such modeling requires fitting of structural 
parameters to ACS multidrizzled data and is beyond the 
scope of this paper.

\subsubsection{Rotation curve modeling from spectra alone}
\label{sec-rcfit}

In the spirit of obtaining kinematic measurements for as many
galaxies as possible, we developed a simplified method for fitting 
seeing-compensated rotation curves using only the information
contained in the 2-d spectra.  This program, {\sc ROTCURVE},
works by constructing models of the unblurred spatially resolved
emission intensity, velocity and dispersion, blurring them along the 
slit to model seeing, and fitting to the data.  Because it 
only has one dimension of spatial information, the rotation 
model and the seeing convolution are 1-d functions along the
slit, and there is no compensation for disk inclination or
slit position angle.

We chose a subsample of galaxies for rotation curve fitting
using cuts in intensity of the strongest emission line and
in spatial extent along the slit.  The strongest emission
line was required to have integrated intensity $>3000~e^-$/pixel \AA\
in the summed 1-d spectrum.  A gaussian was fit to the light profile
along the slit and we required $\sigma_{light}>0.4\arcsec$,
equivalent to FWHM $>0.94\arcsec$.  464 galaxies were selected
for ROTCURVE fitting.  Of these 445 also had ACS imaging and
ellipticity/position angle data, and 380 of the rotation curve
fits were judged to be good by visual inspection.

{\sc ROTCURVE} fits to a single emission line for each object.  It
first fits an emission line to data in each row of pixels, using a
similar algorithm to {\sc LINEFIT}, to obtain profiles of velocity and
dispersion along the slit, and rejects rows with discrepant values
using automatic criteria, testing for e.g. large row-to-row jumps.  It
measures the light distribution along the slit of the continuum plus
emission, fits a gaussian to this profile, and subtracts the assumed
seeing (here 0.7\arcsec) in quadrature to determine the intrinsic,
unblurred intensity profile, modeled as a gaussian $G(x)$ with 
dispersion $r_I$.  A minimum $r_I=0.2\arcsec$ is imposed to keep
intrinsic profiles of small objects from becoming pointlike.

About half the TKRS masks have seeing $=0.7\arcsec$ and the 
remainder have seeing up to 1.0\arcsec, determined from the
spectra of identified stars and the distribution of fitted widths
of galaxies.  Assuming a smaller seeing than the actual leads
to a small underestimate of \vrot.  Assuming a larger 
seeing than the actual causes the {\sc ROTCURVE} modeling to fail on
small objects, because it derives a too-small intrinsic 
spatial extent $r_I$, and the rotation velocity is poorly
constrained due to the assumed large seeing blur.

{\sc ROTCURVE} then constructs models of the intrinsic position-velocity
distribution along the slit.  The velocity model is an arctangent
rotation curve centered at the peak of continuum+emission light,
with a spatially-constant velocity dispersion, so that the
intensity in position-velocity space before blurring by seeing
and instrumental resolution is:

\begin{equation}
I(x,v) = G(x) {\rm exp}(-\frac{(v-V(x))^2}{2\sigtwod^2}),\\
\end{equation}

\begin{equation}
G(x) = \frac{I_{tot}}{\sqrt{2\pi} r_I} {\rm exp}(-\frac{(x-x_0)^2}{2r_I^2}),\\
\end{equation}

\begin{equation}
V(x) = V_{rot} \frac{2}{\pi} ~ {\rm arctan} (x/r_v).
\end{equation}

$G(x)$ is the intrinsic gaussian light distribution along the slit.  
$V(x)$ is the rotation curve with asymptotic circular
velocity \vrot\ and knee radius $r_v$, and  \sigtwod\ is the velocity
dispersion, assumed constant along the slit.  For the case of
the \oii\ 3727 doublet, we use a double gaussian for the distribution 
in velocity, with intensity ratio 1.4.  

The unblurred $I(x,v)$ is then blurred for seeing with a 
1-d gaussian in the spatial direction.  The seeing has the 
effect of mixing gas at different velocities together,
smoothing out velocity gradients and increasing the observed 
velocity dispersion at gradients (see Section \ref{sec-blurdisk}
for an illustration
of this effect).  {\sc ROTCURVE} then takes
moments of the blurred $I_{blur}(x,v)$ in the velocity direction 
to produce a model rotation curve and dispersion profile,
and computes $\chi^2$ of the data--model.  It also fits for an
offset in velocity since the published redshifts can vary from the 
systemic velocity, especially when there are large velocity gradients. 

The model parameters that can be varied are \vrot, \sigtwod, and $r_v$.
For each galaxy, we construct a grid of models over \vrot\ and 
\sigtwod.  Because the models are fairly simple and
involve a series of 1-d convolutions, a brute-force 
minimization is practical; the minimum in $\chi^2$ on the grid
is always unambiguous and well localized.  The best-fit values
of \sigtwod\ are near-quantized in units of the grid spacing 
(here 5 \kms) due to the minimum-finding technique we used.



\begin{figure*}[ht]
\begin{center}
\includegraphics[angle=-90,width=6.5truein]{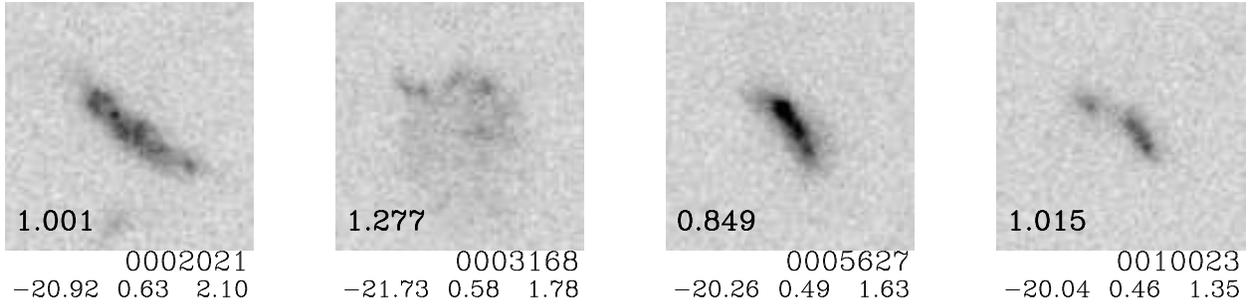}
\caption{ACS $I$-band images for the example galaxies whose
rotation and dispersion are shown in Figure \ref{fig-examplerc}:
TKRS 2021, 3168, 5627, and 10023.  The images are 3\arcsec\ on a side;
north is up and west is to the right.
The panels are labeled with redshift (in the image), TKRS ID, restframe
$M_B$, $U-B$, and linewidth \logsigoned.  The DEIMOS slit PAs are
51\mydeg, 14\mydeg, 37\mydeg, and 39\mydeg\ respectively, north through
east.  The galaxies were classified as edge-on, LSB/irregular, chain, and
hyphen/chain morphology, respectively.
}
\label{fig-exampleimages}
\end{center}
\end{figure*}


\begin{figure*}[ht]
\begin{center}
\includegraphics[angle=-90,width=6.0truein]{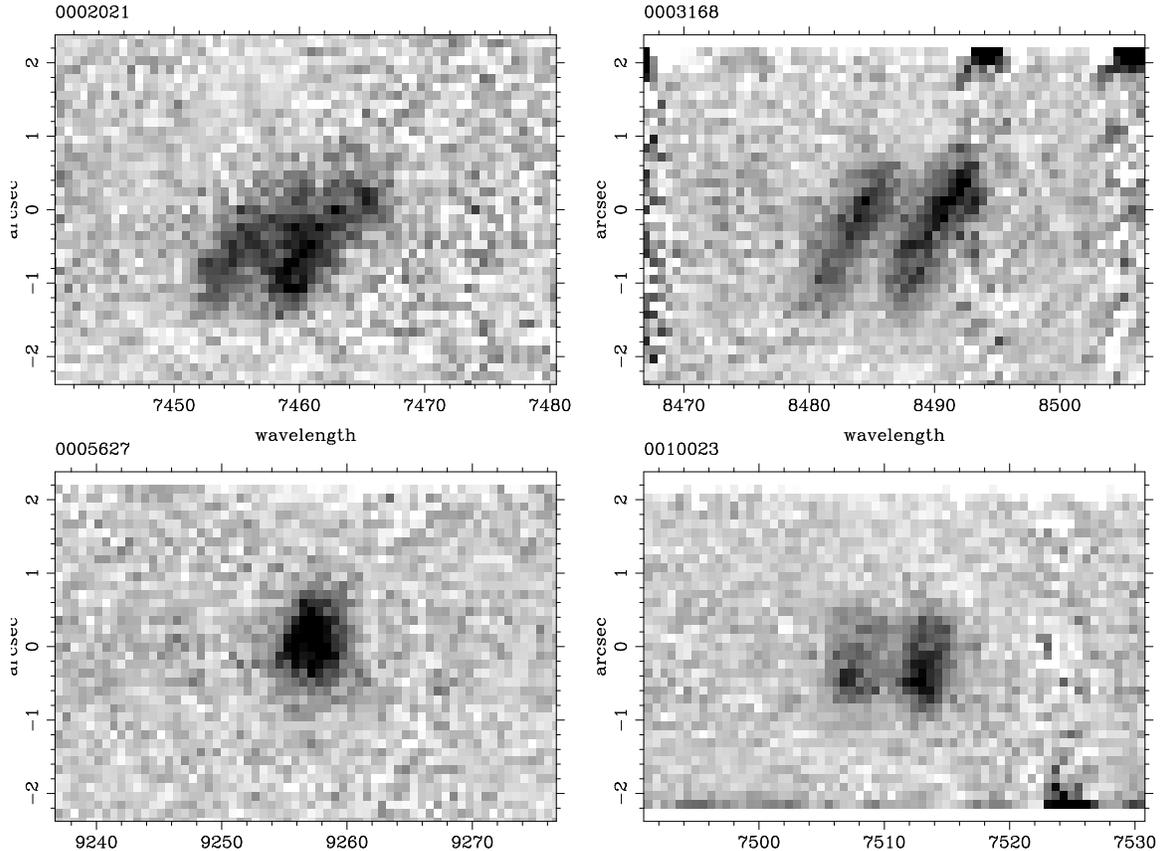}
\caption{Postage stamps of the DEIMOS spectra for the example galaxies 
whose rotation and dispersion are shown in Figure \ref{fig-examplerc}:
TKRS 2021, 3168, 5627, and 10023.  The spectra have been sky-subtracted,
and rectified for display purposes so that columns are constant wavelength.
The galaxy continua, at low S/N/pixel, are not visible in these 
postage stamps.  Each image shows the emission
line used for {\sc ROTCURVE} fitting, the \oii\ 3727 doublet for 
2021, 3168, and 10023, and \oiii\ 5007 for 5627.  TKRS 2021 and 3168
show tilted emission lines, the signature of rotation curves,
but TKRS 5627 and 10023 show very little velocity gradient.
}
\label{fig-examplespec}
\end{center}
\end{figure*}


The rotation scale length $r_v$, describing how fast the rotation curve 
rises, is not strongly constrained by the {\sc ROTCURVE} approach due 
to the seeing blur and the lack of 2-d modeling and position angle 
information.  We chose a fixed $r_v=0.2\arcsec$ for all galaxies, 
which can fit most rotation profiles; 0.1\arcsec\ and 0.3\arcsec\ fit
worse for most galaxies.  The knee radius and circular velocity are 
moderately covariant because the rotation curve turnover is not well 
resolved; changing $r_v$ from 0.2\arcsec\ to 0.1\arcsec or 0.3 \arcsec\
changes \vrot\ by about 0.1 dex in the mean; larger $r_v$ causes 
larger \vrot.  The effect on \vrot\ caused by assuming 0.5\arcsec\ or
0.9\arcsec\ seeing is about 0.05 dex.  The
combination of \vrot\ and \sigtwod\ in quadrature, discussed 
in Section \ref{sec-rotdispsigmacomp} below, 
changes by only half as much as \vrot.

\begin{figure*}[ht]
\begin{center}
\includegraphics[width=6.0truein]{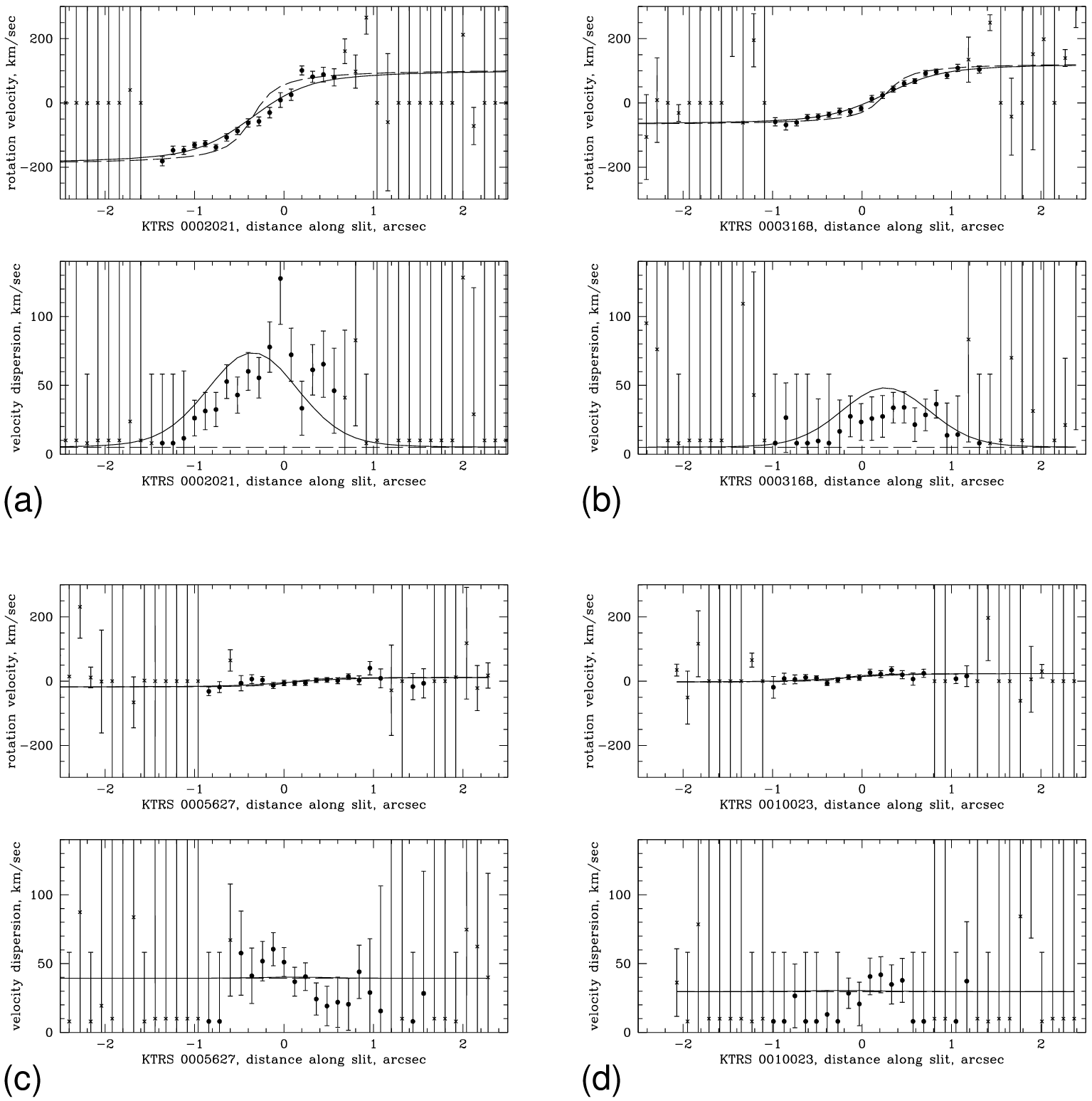}
\caption{Rotation and dispersion profiles for four example galaxies
with well aligned slits.
TKRS 2021 and 3168 at $z=1.00$ and 1.28 are rotation dominated; 
TKRS 5627 and 10023 at $z=0.85$ and 1.02
are dispersion dominated.  The slit PA offsets are 6\mydeg, 9\mydeg, 
3\mydeg, and 10\mydeg.
The points are the observed rotation and dispersion along the slit; 
filled circles 
are the data used by {\sc ROTCURVE} while crosses were rejected by
the program.  The dashed curves show the intrinsic rotation and
dispersion model profiles, and solid curves are the models after the
seeing blur is applied.  In panels (c) and (d) the solid curve
overlies the dashed curve.  The fitted rotation velocities and
dispersions are
$\vrot = 149, 97, 16, 14$ \kms\ and $\sigtwod = 5,5,39,30$ \kms\
respectively.
}
\label{fig-examplerc}
\end{center}
\end{figure*}

Figure \ref{fig-exampleimages} shows ACS $I$-band images
of four example galaxies at $z \sim 1$ from the TKRS.
Figure \ref{fig-examplespec} shows postage stamp images
of their emission line spectra, and
Figure \ref{fig-examplerc} shows their rotation and
dispersion profiles fit by {\sc ROTCURVE}.  In Figure
\ref{fig-examplerc}, the points are 
velocity and dispersion from the spectra; the dashed and solid
lines show the best-fit model before and after applying the 
seeing blur.  These galaxies are relatively clean examples
of a dichotomy in kinematic profiles that is common in the TKRS data.

TKRS 2021 and 3168 show the familiar shape of a rotation 
curve.  Beam smearing by the seeing not only smooths
out the gradient in velocity, but induces a peak in the 
observed dispersion; the intrinsic dispersion is significantly
less than the observed.  TKRS 5627 and 10023 are less familiar. 
They have quite small rotation, but require a velocity dispersion 
component to fit the data, with $\sigtwod>\vrot$.  
Galaxies like TKRS 5627 and 10023 are
common in our sample and are the reason we formulated {\sc ROTCURVE}
to include the velocity dispersion \sigtwod\ as a second free
parameter in the modeling.  Note that these galaxies are relatively 
small but not extreme starbursts; they are not compact, nor
very faint.  We discuss rotation and dispersion
dominated galaxies further in Sections \ref{sec-rotdispdom}
and \ref{sec-rotdispnature} below.


We rarely see a rotation curve in which the observed points
clearly roll over onto the flat part of the rotation curve,
especially at $z \gtrsim 0.4$.
This is partly due to the limited extent of the data, but 
frequently a more important effect is the seeing blur,
which makes the observed rotation curve rise shallower
than the true rise.  The effect seen in TKRS 3168 in
Figure \ref{fig-examplerc} is common: the deblurred model rotation
curve has neared its asymptotic value while the observed
data are still rising.  Despite the lack of a flat-part
in the observed data, the fitted values faithfully 
represent the observed velocity spread.  Objects in
which the data covered only a small spatial extent and
the fitted velocity extrapolated beyond the data were
rejected as erroneous fits.

Because the rotation curve data points are highly correlated,
$\chi^2$ does not yield errors on \vrot\ and \sigtwod\
directly.  The fitted velocity \vrot\ is a good representation
of the gradient seen in the rotation curve, but subtler details 
such as the true errors on \vrot\  and its dependence on the 
spatial extent of the data will require further Monte Carlo 
simulation.  Here and in Paper II, measuring Tully-Fisher
evolution with \vrot\  and \sigtwod\ is secondary to using them to
probe the nature of galaxy kinematics, and as a test of the
meaning of integrated linewidth \sigoned.

As a check on the fits and models, 
{\sc ROTCURVE} uses the observed row-by-row-fit 
rotation velocity and dispersion profile to reconstruct the 2-d data, 
and subtracts it from the original data to produce a residual map.  
It also produces reconstructed 2-d data from the best-fit blurred 
model (convolving back to the instrumental resolution).  
The reconstruction from the row-by-row fits is generally very
good.  The reconstruction from the model is also generally good, 
although it leaves residuals in cases such as an asymmetry in
emission intensity, since the model is symmetric by construction.
In principle, the models could be fit directly to the 2-d data
without ever reducing to a 1-d rotation curve, but the
rotation curve is a useful tool for evaluating the data and model.

The {\sc ROTCURVE} fitting procedure runs automatically.
Afterwards we inspected the rotation and
dispersion profiles to reject bad or erroneous fits.  380 of
the 445 galaxies passed this inspection.  The dominant causes of 
failures are: discrepant points in the rotation curve which 
evade {\sc ROTCURVE}'s automatic rejection; asymmetry which
offsets the photometric center from the kinematic center; or
small spatial extent of the rotation curve: we rejected most
objects for which the usable rotation curve data spanned 
less than 1\arcsec.

{\sc ROTCURVE} departs from previous high-redshift rotation curve
modeling by making the velocity dispersion \sigtwod\ a free parameter; 
to our knowledge, other models have all fixed the dispersion to
some low value.  Fixing the dispersion is probably acceptable
for samples of only morphologically-normal inclined disk galaxies
observed with well-aligned slits, where rotation is the chief kinematic
support and is unambiguously detected.  
However, as seen for TKRS 5627 and 10023 and shown in Section
\ref{sec-rotdispdom}, there are galaxies with low rotation and
large dispersion in our broader sample, and for these, fitting
without a dispersion term will yield erroneous rotation 
velocities.  

The dispersion term \sigtwod\ in {\sc ROTCURVE}
does not have to correspond to a literal gas velocity dispersion
similar to the velocity dispersion of stars in an spheroidal
galaxy.  Because we are measuring nebular emission lines that
occur in gas at $T \sim 10^4$ K, the gas in any individual \hii\ 
region will have a dispersion of only 8-10 \kms.  Taken literally,
a dispersion $\sigtwod \sim 30$ \kms implies a gas temperature 
of $T \sim 10^5$ K, which is an unfavorable place on the ISM cooling 
curve.  It is more likely that the dispersion comes from relative 
velocities of discrete \hii\ regions.  

Values of $\sigtwod \gtrsim 20$ \kms\ can represent an effective 
dispersion, caused by the blurring of velocity gradients on 
scales at or below the seeing limit.  Such a seeing-induced
dispersion \sigtwod\ can appear for slits which are aligned
close to the minor axis of the galaxy and miss the resolved
velocity gradient, as shown in Section \ref{sec-blurdisk}.  
But for TKRS 5627 and 10023, and similar galaxies in our
sample, the slit was well aligned with the photometric
major axis, yet there is almost no ordered velocity gradient.
The effective dispersion \sigtwod\ must be explained by 
non-ordered or disturbed gas kinematics on angular scales well below
the seeing blur; these motions may not even have a preferred plane.
The geometry of the \hii\ regions, and why they have not dissipated 
into a rotating disk, cannot be measured with these seeing-limited 
observations.

\subsection{Photometry}

Nearly all TKRS galaxies have $BViz$ photometric measurements from the
GOODS ACS catalog (Giavalisco \etal\ 2004).  Ground-based photometry in
$UBVRIz^\prime(HK^\prime)$ for a large field including GOODS-N is
available from Capak \etal\ (2004).  We used the 2.8\arcsec\ diameter
aperture magnitudes from the GOODS catalog and the 3\arcsec\ diameter
magnitudes from the Capak catalog.  For all but the largest,
lowest-redshift galaxies, aperture corrections are small, so we omit
them.  Of the 1440 TKRS galaxies, 1423 have matches with 
optical magnitudes in the Capak \etal\ catalog.  Fourteen
of the remainder have magnitudes from the GOODS ACS catalog,
and only 3 have neither.  

The Capak \etal\ $HK^\prime$ data are shallower than the optical data 
and the depth is not constant over the GOODS field.
We used only aperture $HK^\prime$ magnitudes that have errors 
$<0.3$ mag, yielding $HK^\prime$ for 919 TKRS galaxies.  With this
error cut, the fraction of TKRS galaxies with $HK^\prime$ magnitudes
reaches 50\% at $HK^\prime = 22.3$ (AB).

Our primary interests are to measure consistent restframe
absolute blue magnitude $M_B$ and color $U-B$, and a rest infrared
magnitude $M_J$ and $R-J$ color.  We convert to rest $B$ and $J$
because the observed filters cover or come relatively close to these
restframe bands over $0<z<1.5$, minimizing extrapolation.
We used the
template-fitting $K$-correction procedure described in 
previous DEEP papers (Weiner \etal\ 2005; Willmer \etal\ 2006)
to convert observed colors into $K$-correction and restframe
color.  Briefly, this method uses redshifted templates:
34 spectra of local galaxies with wavelength coverage 1200\AA\
 -- 1 $\mu$m (Kinney \etal\ 1996) for $M_B$ and $U-B$; and 9 templates
generated by PEGASE (Fioc \& Rocca-Volmerange 1997) representing
E, Sb and Sd SEDs at 1,5, and 10 Gyr for $M_J$ and $R-J$.
To analyze, for example,
a galaxy at $z=0.8$ with $R-I=1.2$, the templates are redshifted 
to $z=0.8$ and their observed $R-I$ colors synthesized.  Since
the templates' restframe $U-B$ colors and $K$-corrections from $I$ 
to $B$ are known, we fit a low order polynomial to $U-B$ as
a function of $R-I$ and evaluate this fit at $R-I=1.2$ to 
find $U-B$ for the galaxy under analysis.

This procedure works best when the observed filter pair
transforms relatively closely to the wavelength of the restframe 
filters, as is the case for $R-I$ at $z=0.8$ into rest $U-B$.  
When the transformation is not close, there is larger scatter
about the fit of $U-B$ as a function of observed color, so choosing
an appropriate filter pair is important.  

To derive the restframe $U-B$ and $M_B$ presented here,
we used the filter pairs of Capak $B-V$,$R-I$, and $I-z$,
switching over from one pair to the next at $z= 0.6,1.1$ respectively.
For $\Delta z=0.1$ around the switchover redshift we interpolate 
to make a smooth transition.  For rest $R-J$ and $M_J$ we use the 
observed $I-HK^\prime$ filter pair and the $HK^\prime$ magnitude.

In practice, the $K$-corrections are fairly stable.  However, certain
filter pairs produce discordant results for restframe colors:
e.g. ACS $B-V$ and $V-i$ produce colors for galaxies at $z=0.5$
that are $\sim 0.1$ mag offset.  This problem is most obvious
with the ACS filter pairs and with the Capak $V-R$ color.  
In general, a warning sign of the problem shows up as template mismatch
in a color-color plot: for certain combinations of redshift and 
color pairs, galaxies and redshifted templates do not fall on 
exactly the same locus.  The reasons for these mismatches are
not clear; they may be a combination of small zeropoint
shifts and imperfect filter/instrument response curves.
The effect is small on absolute magnitudes, and relative color
measurements at a given redshift are reliable, but 
measurements of absolute color evolution over a large range
in redshift can be affected.

We found that using the Capak $B-V$, $R-I$, $I-z$ color sets
minimizes the effect of these mismatches; using the ACS filters
produces slightly abnormal trends in restframe color with
redshift.  Comparing the absolute 
magnitudes and colors from the Capak color sets and from the fully
independent ACS $B-V$, $V-i$, $i-z$ sets, we find that 95\%
of the objects have a magnitude difference $<0.33$ and a
color difference $<0.26$, yielding $1\sigma$ errors 
of 0.12 in $M_B$ and 0.09 in $U-B$.
The errors on observed optical magnitudes from Capak \etal\ (2004)
are 0.05-0.07 mag at the TKRS limit; additional scatter is
propagated into the absolute magnitudes by the observed color
error and the template scatter in the K-correction procedure.

The median error on $HK^\prime$ magnitude for galaxies with 
kinematic measurements is 0.14 mag.  Although we are using
model rather than empirical SEDs to K-correct the IR 
magnitude, galaxy SEDs have less variety in the IR than the
optical and so the K-correction depends only weakly on the
specific templates used.  The error on the IR observations 
dominates over the IR template scatter.

\section{Sample magnitude, color and completeness}
\label{sec-complete}

Of the 1440 TKRS galaxies, 1423 have magnitudes from Capak 
\etal\ (2004) and 14 have magnitudes from ACS (Giavalisco \etal\ 2004).
893 galaxies have magnitudes and a linewidth
determination, and an 
additional 196 are kinematically unresolved.  Of these 1089,
913 are brighter than $M_B=-18$, a limit we impose
for Tully-Fisher fitting in Paper II.
681 of the 1089 with kinematics have a reliable $M_J$ and
647 of these are brighter than $M_J = -19$.
Table \ref{table-catalog} lists TKRS galaxies, positions, redshifts,
observed magnitudes, restframe magnitudes and colors, and linewidths. 

\begin{deluxetable*}{rrrrrrrrrrrrrrrr}

\tablecaption{
Catalog of TKRS galaxies and linewidth measurements
\label{table-catalog}
}

\tablecolumns{16}
\tablewidth{0pt}
\tabletypesize{\scriptsize }

\setlength{\tabcolsep}{0.0in}

\tablehead{
 TKRS  & Capak & RA & Dec & $z$\tablenotemark{1} & 
   $B$\tablenotemark{2} & $R$\tablenotemark{2} & $I$\tablenotemark{2} & 
   $HK^\prime$\tablenotemark{2} & error\tablenotemark{2} &
   $M_B$\tablenotemark{3} & $U-B$\tablenotemark{3} & 
   $M_J$\tablenotemark{3} & $R-J$\tablenotemark{3} & 
   log~$\sigoned$\tablenotemark{4} & error\tablenotemark{4}
 \\
 ID  &  ID  & (2000) & (2000) & & & & & & ($HK^\prime$) & & & & & & (log~$\sigoned$)
 }
 \startdata
0000034 & -99.9 & 189.019492 & 62.261711 & 0.4581 & -99.90 & -99.90 & -99.90 & -99.90 & -99.90 & -99.99 & -99.99 & -99.90 & -99.90 & 1.648 & 0.173 \\
0000117 & 59102 & 189.232904 & 62.360603 & 0.5852 & 24.71 & 23.66 & 23.34 & 22.33 & 0.36 & -18.46 & 0.47 & -19.88 & 0.69 & 1.483 & 0.699 \\
0000225 & 62458 & 189.178462 & 62.333772 & 0.8547 & 24.49 & 24.09 & 24.08 & -99.90 & -99.90 & -19.08 & 0.07 & -99.90 & -99.90 & 1.621 & 0.068 \\
0000239 & 45551 & 189.084525 & 62.291556 & 0.8480 & 25.10 & 23.66 & 22.70 & 20.97 & 0.11 & -20.37 & 0.90 & -22.27 & 0.91 & 1.879 & 0.084 \\
0000241 & 62192 & 189.197325 & 62.342981 & 0.8894 & 24.29 & 23.84 & 23.12 & -99.90 & -99.90 & -20.12 & 0.68 & -99.90 & -99.90 & 1.357 & 0.151 \\
0000276 & 58390 & 189.292225 & 62.386944 & 1.2590 & 24.23 & 23.56 & 23.15 & -99.90 & -99.90 & -21.58 & 0.77 & -99.90 & -99.90 & -99.900 & -99.900 \\
0000320 & 63030 & 189.136354 & 62.312436 & 0.4826 & 24.81 & 24.11 & 23.88 & -24.14 & 0.87 & -17.41 & 0.29 & -99.90 & -99.90 & 1.400 & 0.200 \\
0000345 & 50539 & 188.929450 & 62.215525 & 0.8509 & 24.41 & 23.66 & 22.97 & 21.58 & 0.37 & -20.10 & 0.65 & -21.64 & 0.80 & -99.900 & -99.900 \\
0000428 & 58669 & 189.274379 & 62.375206 & 0.4870 & 24.31 & 23.78 & 23.53 & -24.26 & 2.01 & -17.87 & 0.26 & -99.90 & -99.90 & 1.400 & 0.200 \\
0000432 & 62868 & 189.150483 & 62.318058 & 0.6809 & 23.83 & 22.48 & 21.88 & 20.87 & 0.06 & -20.44 & 0.65 & -21.72 & 0.66 & 1.874 & 0.037 \\
0000444 & 46466 & 189.006417 & 62.253725 & 0.2985 & 23.44 & 22.07 & 21.72 & 21.15 & 0.19 & -18.19 & 0.74 & -19.44 & 0.45 & 1.499 & 0.150 \\
0000445 & 62021 & 189.214375 & 62.348203 & 0.7441 & 23.58 & 22.77 & 22.25 & 21.45 & 0.22 & -20.40 & 0.53 & -21.33 & 0.46 & 1.650 & 0.129 \\
0000448 & 59290 & 189.223667 & 62.353142 & 0.4724 & 23.14 & 22.08 & 21.83 & 20.98 & 0.14 & -19.54 & 0.58 & -20.71 & 0.60 & 1.693 & 0.121 \\
0000463 & 62992 & 189.143487 & 62.313794 & 0.5027 & 24.66 & 23.87 & 23.66 & 22.56 & 0.24 & -17.67 & 0.31 & -19.29 & 0.77 & 1.579 & 0.149 \\
0000484 & -99.9 & 189.171067 & 62.326453 & 0.6088 & 23.96 & -99.90 & 23.06 & -99.90 & -99.90 & -19.07 & 0.39 & -99.90 & -99.90 & 1.400 & 0.200 \\
0000555 & 46181 & 189.033342 & 62.264919 & 0.4589 & 23.21 & 21.88 & 21.58 & 20.89 & 0.24 & -19.66 & 0.73 & -20.72 & 0.47 & 1.885 & 0.073 \\
0000584 & 46681 & 188.998854 & 62.246286 & 0.5306 & 23.85 & 22.55 & 22.23 & 21.31 & 0.21 & -19.23 & 0.61 & -20.67 & 0.64 & 1.400 & 0.200 \\
0000593 & 58521 & 189.284400 & 62.381100 & 0.2046 & 24.31 & 23.66 & 23.51 & 23.44 & 0.99 & -15.86 & 0.36 & -16.33 & 0.07 & 1.400 & 0.200 \\
0000617 & 46339 & 189.025129 & 62.258847 & 0.4542 & 23.68 & 22.58 & 22.41 & 22.18 & 0.38 & -18.90 & 0.58 & -19.41 & 0.17 & 1.274 & 1.467 \\
0000618 & 58908 & 189.266258 & 62.368506 & 1.0080 & 24.36 & 23.92 & 23.18 & 22.64 & 0.73 & -20.59 & 0.78 & -20.81 & 0.19 & 1.488 & 0.289 \\

 \enddata


\tablecomments{The complete version of this table is in the electronic edition 
of the Journal.  The printed edition contains only a sample.  Undefined values
are given as $<-99$.}

\tablenotetext{1}{Redshift.}
\tablenotetext{2}{Apparent $BRI(HK^\prime)$ AB aperture magnitudes (Capak 
\etal\ 2004).  Galaxies with only $BI$ use GOODS ACS aperture magnitudes
(Giavalisco \etal\ 2004).}
\tablenotetext{3}{Restframe $M_B$, $U-B$, $M_J$ and $R-J$ AB magnitudes
from the SED $K$ correction procedure described in text.}
\tablenotetext{4}{\,Log linewidth in \kms\ from integrated emission and 
its error.  Galaxies assigned a log linewidth of 1.4 and error of 0.2 are 
kinematically unresolved, see text.}

\end{deluxetable*}


\begin{figure*}[ht]
\begin{center}
\includegraphics[width=5.5truein]{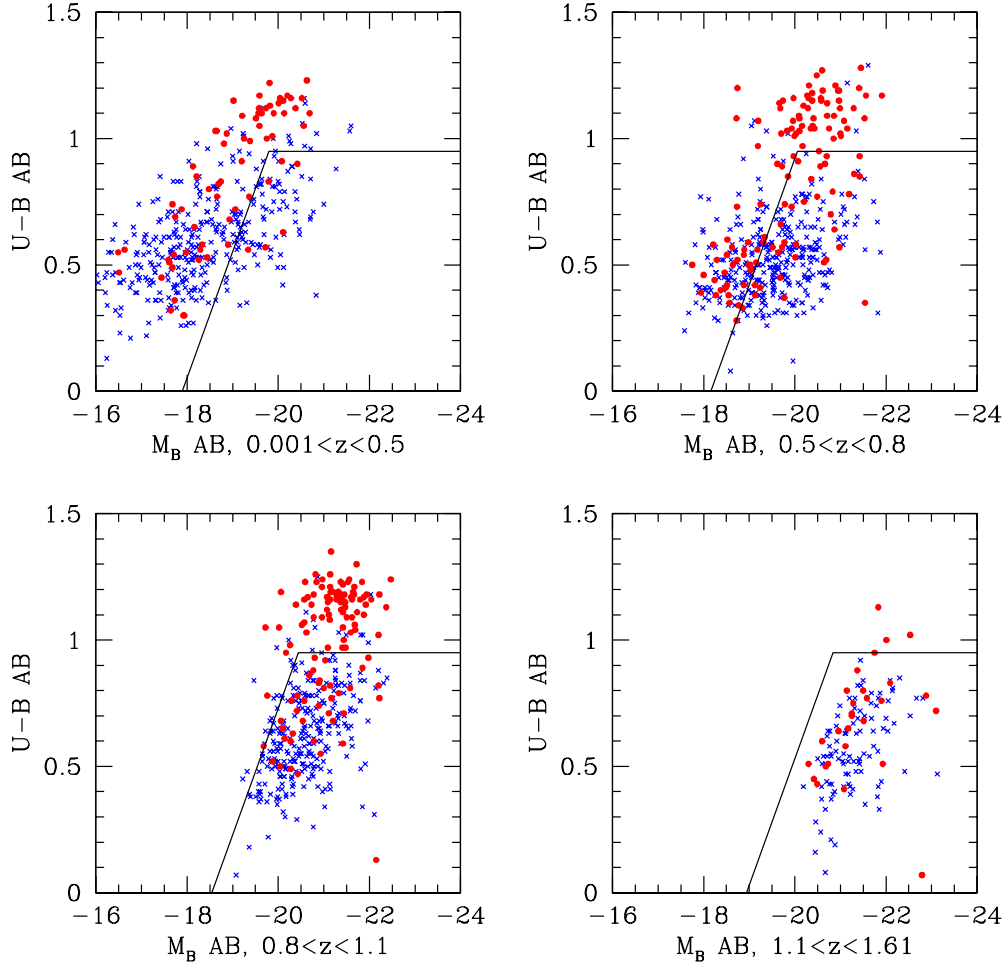}
\caption{Absolute magnitude and color for TKRS galaxies in 4 redshift
ranges.  Blue Xes: galaxies with good linewidths; red circles: galaxies 
with poorly measured linewidth.  The bulk of linewidth failures are in the
red side of the color bimodality.
The horizontal and diagonal lines show a magnitude and color selection
that constructs samples whose depth is matched to the selection limit
at $z=0.9$ and an evolving $L^*$.
}
\label{fig-magcolor}
\end{center}
\end{figure*}


Figure \ref{fig-magcolor} shows the restframe optical color-magnitude 
distribution of TKRS galaxies and the location of the 1089 galaxies
with and 348 without measured linewidths.  The TKRS galaxies exhibit
the color bimodality now well-established locally and to $z\sim 1$
(Strateva \etal\ 2001; Bell \etal\ 2004; Wirth \etal\ 2004; 
Weiner \etal\ 2005).
Both red and blue galaxies show a shallow color-magnitude relation:
brighter galaxies are redder in the mean.
The majority of galaxies for which we fail to get a 
good linewidth are on the red side of the bimodality, $U-B>0.95$, 
because those galaxies have weak or
no emission lines (e.g. Weiner \etal\ 2005).  Among the
blue galaxies, linewidth failures are rare and fairly randomly
distributed.  We examined each failure and found that most are 
due to low line flux or night sky residuals.  The failure rate in
blue galaxies is small and there is no evidence
that failures occur preferentially at large or small dispersions.

At low redshift, blue galaxies in the TKRS sample show the well-known
color-magnitude relation; fainter galaxies are bluer (upper left
panel of Figure \ref{fig-magcolor}).  At higher redshift,
the faint limit is due to the TKRS magnitude selection;
the selection limit in apparent $R$ magnitude corresponds
to a tilted line in the restframe $M_B, U-B$ plane (e.g.\ lower
left panel of Figure \ref{fig-magcolor}).  The tilt of 
this line changes with redshift.  In Paper II we test the effect
of the selection limit on Tully-Fisher relations by defining
an approximately mass-matched sample, applying the $z \sim 0.9$ tilt
at all redshifts and a magnitude cut that evolves similarly
to $L^*$.  This cut is shown by the diagonal lines in Figure
\ref{fig-magcolor}.


\begin{figure}[ht]
\begin{center}
\includegraphics[width=3.5truein]{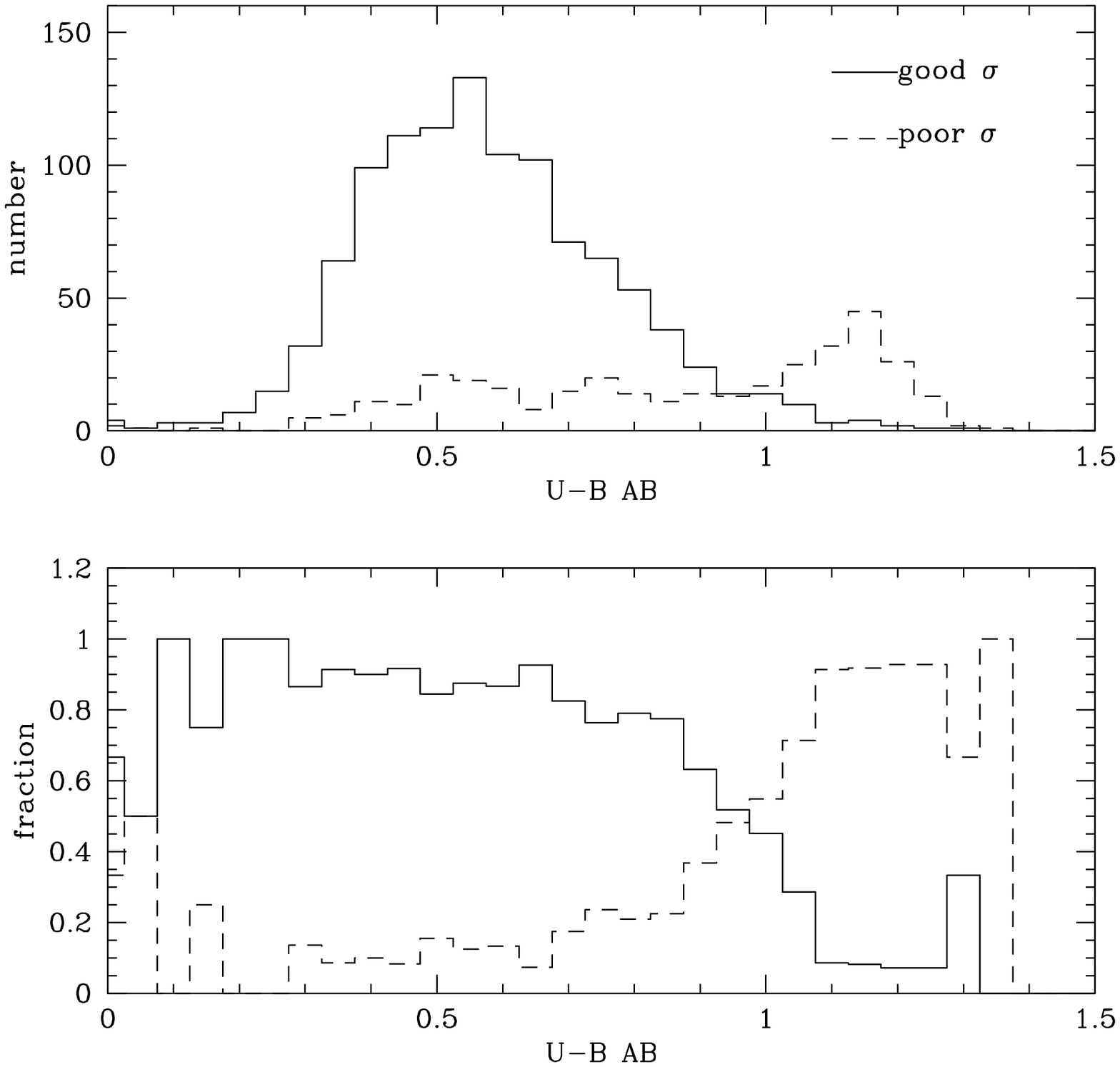}
\caption{Linewidth completeness as a function of restframe $U-B$ color.
Solid line: galaxies with good linewidths; dashed line: galaxies
with poorly measured linewidth.  The top panel shows absolute numbers
and the bottom panel shows good or bad as a fraction of the total
with redshifts.}
\label{fig-ubcomplete}
\end{center}
\end{figure}


The redshift completeness of the TKRS is $\sim 70\%$ to $R_{AB}=24.4$,
discussed by Wirth \etal (2004).  Redshift failures in the TKRS are 
mostly faint blue galaxies that are probably at $z>1.5$.  Here we
discuss the linewidth completeness, i.e. the fraction of galaxies
with good linewidths over galaxies with redshifts.

Figure \ref{fig-ubcomplete} shows the linewidth completeness as
a function of rest $U-B$ color.  The completeness is roughly
constant for blue galaxies but drops sharply for red galaxies.
Because red galaxies are bright, a histogram of completeness 
as a function of rest $M_B$ for the whole sample actually
drops at bright magnitudes.  


\begin{figure}[ht]
\begin{center}
\includegraphics[width=3.5truein]{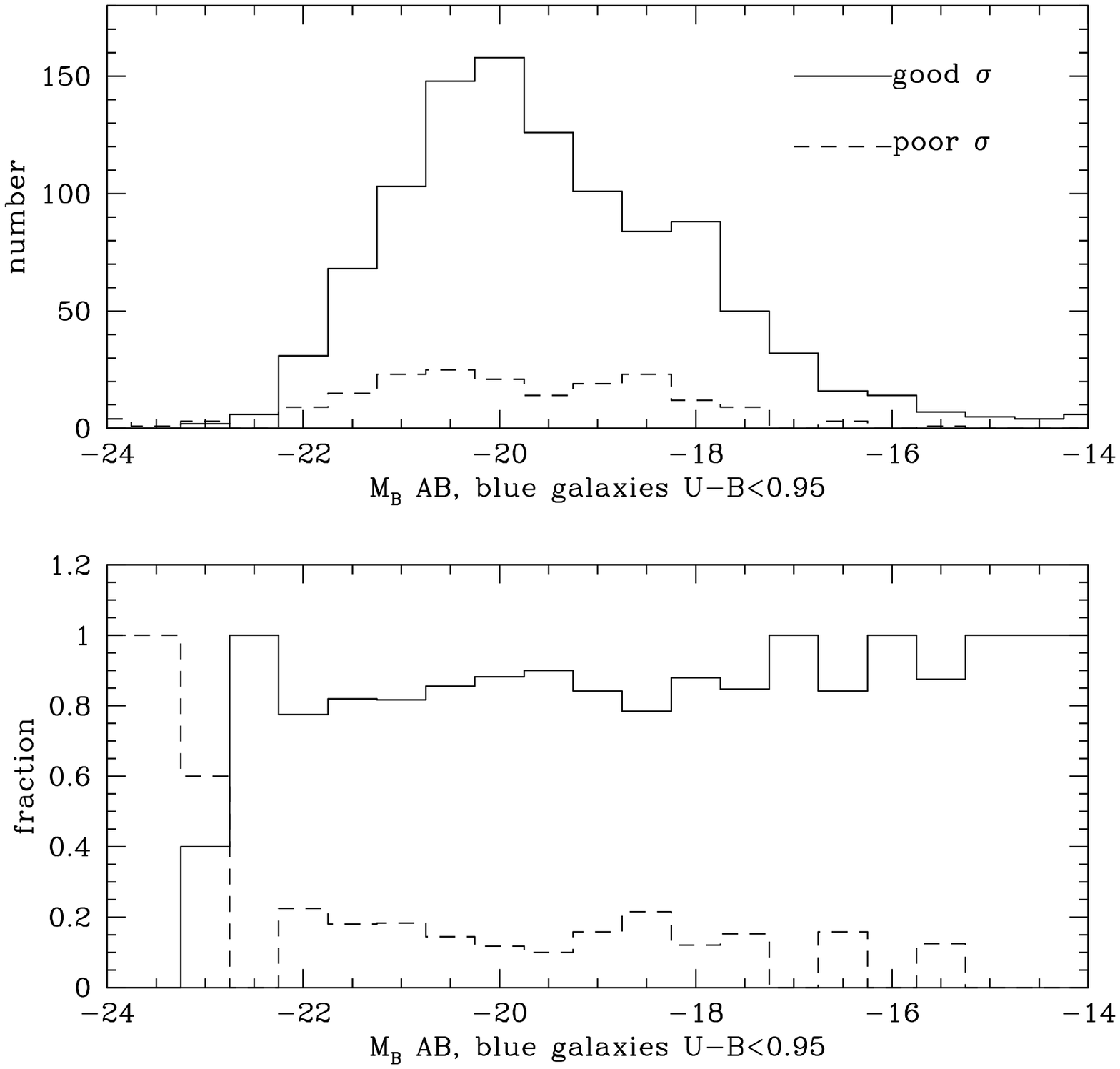}
\caption{Linewidth completeness for blue galaxies
as a function of absolute magnitude $M_B$.
Solid line: galaxies with good linewidths; dashed line: galaxies
with poorly measured linewidth.  The top panel shows absolute numbers
and the bottom panel shows good or bad as a fraction of the total
of blue galaxies with redshifts.}
\label{fig-mbcomplete}
\end{center}
\end{figure}

\begin{figure}[ht]
\begin{center}
\includegraphics[width=3.5truein]{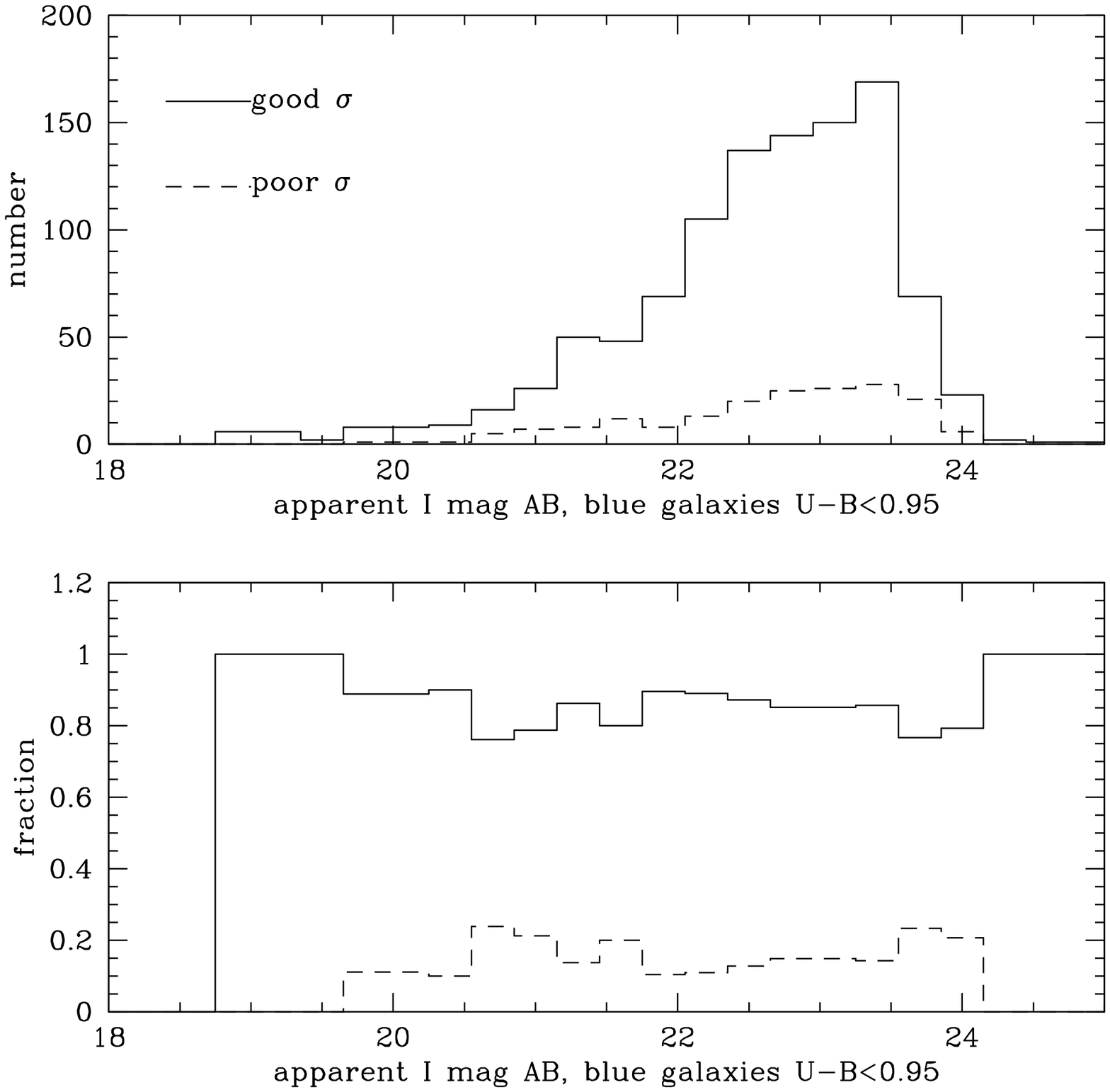}
\caption{Linewidth completeness for blue galaxies
as a function of apparent magnitude $I_{AB}$.
Solid line: galaxies with good linewidth; dashed line: galaxies
with poor linewidth measure.  The top panel shows absolute numbers
and the bottom panel shows good or bad as a fraction of the total
of blue galaxies with redshifts.}
\label{fig-imagcomplete}
\end{center}
\end{figure}


Red galaxies and blue galaxies are two distinct populations and
the bulk of our measurements are blue galaxies, so in this paper
we are constraining the behavior of the blue galaxy population.
Figures \ref{fig-mbcomplete} and \ref{fig-imagcomplete} show
the linewidth completeness of the sample restricted to
TKRS galaxies on the blue side of the color bimodality, $U-B<0.95$.
Linewidth completeness is nearly constant with either
absolute or apparent magnitude, with a small tendency to
be lower for galaxies bright in absolute magnitude (lower
emission EW) or faint in apparent magnitude (lower apparent flux).
The lack of any strong dependence on magnitude reflects the
fact that intrinsically-faint galaxies tend to have high
emission equivalent width.

Because completeness of the linewidth sample depends strongly only 
on color, we have a fair sample of the blue galaxy population,
and our measurements will not be strongly affected by selection
effects.  Of course, any evolution measurement or interpretation
will apply only to blue galaxies.

\section{Properties of kinematic measures}
\label{sec-kinprops}

The limited spatial resolution of high-redshift spectra makes
the interpretation of kinematic measures less obvious than in local
galaxies.  Our goal in this section is to establish the properties
of our sample and determine the effect of observational limitations
on the measures of dispersion and rotation.  

There has been some controversy over how well 
measures such as integrated linewidth probe galaxy kinematics.
Modeling simulated observations of disk velocity fields with
true circular velocity $V_c$,
Rix \etal\ (1997) found that for their observational 
parameters, the average integrated linewidth 
$<\sigma>/V_c = 0.6$ with substantial scatter ($\sim 0.15$) 
depending on the unknown position angles and inclinations,
where $<\sigma>$ incorporates an average over inclination but 
$V_c$ is the true value.  For random orientations, $<\sini> = 0.79$,
leaving an average factor of 0.76 for $\sigma/(V_c \sini)$.
From observations of local galaxies, Kobulnicky \& Gebhardt (2000)
found that the \oii\ emission linewidth tracks the \hi\ linewidth $W_{20}$
well for most galaxies, using $\sigma = 0.28 W_{20}$ (both uncorrected
for inclination), but the \oii\ 
linewidth is low in 2/22 extreme cases.   (The \hi\ width $W_{20}$ is
the full-width of the \hi\ profile at the 20\% flux level.)  This 
relation is roughly consistent with the Rix \etal\ relation since 
$W_{20} \simeq 2(V_c \sini + 15~\kms)$ (Tully \& Fouque 1985).

Lehnert \& Heckman (1996) showed a tendency for edge-on IR-luminous
starburst galaxies to have nuclear emission linewidth that is low compared
to the expected velocity dispersion of the full potential; these
galaxies are mostly fairly large and the nuclear emission
does not probe the full potential.
Barton \& van Zee (2001) observed four blue compact
dwarf galaxies and found that  $\sigma(\ha)$ falls
below optical $V_{rot}$ and \hi\ $W_{50}$.  However, their
four galaxies average $\sigma/V_{2.2} = 0.7$, and the two with
21 cm data have $\sigma/W_{50} = 0.27$ (all quantities uncorrected
for inclination).  These are
actually fairly consistent with the $<\sigma>/V_c$ offset of 
Rix \etal\ and the linewidth / \hi\ width factor of Kobulnicky
\& Gebhardt.  Rotation curves can also be affected:
kinematic distortions and truncations could cause
galaxies to have low measured rotation (Barton \etal\ 2001;
Kannappan \& Barton 2004).

Perhaps the most serious discrepancy between optical and 
\hi\ width is found in a sample of 10 local 
blue compact galaxies (BCGs; Pisano \etal\ 2001), 
which show $W_{20}(H\beta)/W_{20}(\hi) = 0.66 \pm 0.16$.
In terms of dispersion this implies $\sigma(H\beta)/W_{20}(\hi) = 0.18$.
These galaxies are a fairly special population with high
emission equivalent widths and small sizes, $r_e=0.6-1.9$ kpc,
smaller than average for intermediate-$z$ BCGs.
They are most similar to NGC 4449, the furthest outlier in the 
sample of Kobulnicky \& Gebhardt (2000), which has a much larger
\hi\ extent than optical; and to the compact
narrow emission-line galaxies (CNELGs, Koo \etal 1995) in their
sizes.  In the sample
of Pisano \etal, the optical linewidths are most affected
when $W_{20}(\hi)<150$ \kms, which corresponds to an unaffected
optical $\sigma<45$ \kms; smallest galaxies can be the most 
seriously affected. 

These types of underestimation of kinematic width may happen
in high-redshift samples.  They might explain some of the
discrepancies between previous TF evolution measurements, as 
Pisano \etal\ (2001) suggest, especially in samples with
high line EW.  However, even at high redshift, the
extreme BCGs or CNELGs are a minority of
blue galaxies (Koo \etal\ 1995).  These galaxies are quite small;
even with 1-2 mag brightening, the Pisano \etal\ (2001) sample 
would be among the faintest of the $z\sim 1$ galaxies in the TKRS 
sample.  The possibility of BCG-induced offsets does argue in
favor of using large samples which are not selected on 
emission line EW, and in which evolution can be measured 
internal to the sample.

Ideal measures of high-redshift kinematics would yield 
2-d maps with high spatial resolution.  This may soon be
possible with adaptive optics and integral field spectroscopy,
but will remain impractical for large samples for some time.
Here we carry out empirical tests on our sample and simulations
of seeing-blurred observations to test the properties of
linewidth measurements.

\subsection{Spatial extent of emission}

It is frequently suggested that high-redshift galaxies could
have enhanced emission from their centers, such as a nuclear
starburst, and that this could produce smaller linewidths
because the emission does not probe the full potential,
or comes from a part of the rotation curve that is rising
(e.g. Lehnert \& Heckman 1996).
Simard \& Pritchet (1998) found that $\sim 25\%$ of their galaxies
had unresolved emission in their CFHT spectra.
Rix \etal\ (1997)
suggested that $\sim 20\%$ of their sample might have nuclear
emission from high-density gas, based on the \oii\ doublet ratio.
However, both of these samples are selected to have high EW or blue 
colors, and are relatively low S/N compared to the TKRS spectra.  
Another possibility is that line emission could be dominated
by one or a few large HII regions and so not probe the full
velocity field of the disk.


\begin{figure}[ht]
\begin{center}
\includegraphics[width=3.5truein]{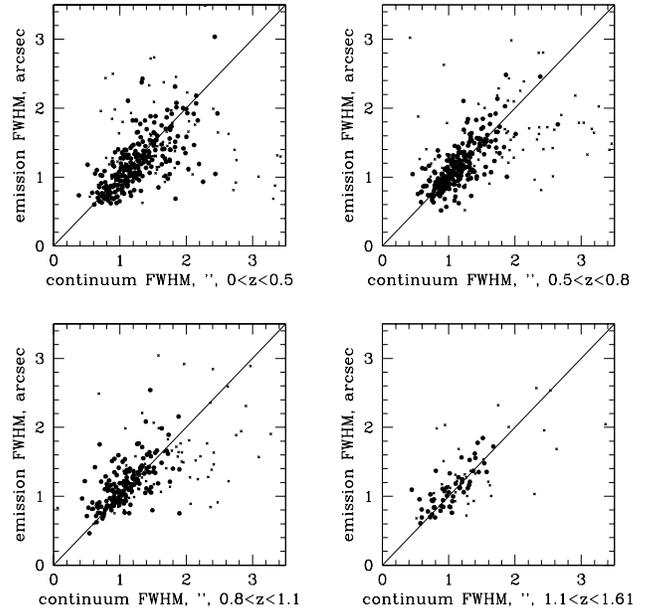}
\caption{Spatial extent along the slit
for TKRS galaxies in four redshift ranges.  
The fitted  gaussian FWHMs of emission and continuum
are plotted.  Filled circles have error on both FWHMs $<0.1''$,
Small Xes have error on one or both FWHMs $>0.1''$, indicating an 
unreliable fit.  The sizes are seeing-limited around
FWHM $\sim 0.7\arcsec$, and above that are consistent with equal sizes 
for emission and continuum.}
\label{fig-emsize}
\end{center}
\end{figure}


We tested the spatial extent of emission in the TKRS linewidth
sample by fitting gaussian profiles along the slit.  We
collapsed the 2-d data in the wavelength direction over
a 15 \AA\ range for emission and two 100 \AA\ ranges on either
side of the line for continuum.  We then fit gaussians
using a non-linear least squares routine.

The fitted spatial widths of emission versus continuum are shown
in Figure \ref{fig-emsize}, divided into four redshift ranges.  
The sizes are seeing-limited around
FWHM $\sim 0.7\arcsec$ and track each other
well at larger sizes.  The small mask-to-mask seeing variations in the
TKRS data will tend to spread galaxies along the 1:1 line, but the
relation continues far beyond FWHM $\sim 1.0\arcsec$.  Outliers from
the 1:1 line generally have a large error in one size measurement.
While this comparison is somewhat limited by seeing, there is
no evidence for a large population of nuclear emission or single-HII-region
emission.  In fact there is strong evidence that emission and continuum 
size track each other well.

\subsection{What integrated linewidths measure: simulation of observations}
\label{sec-blurdisk}

In order to understand what an integrated linewidth is really
measuring, we simulated observations of a disk galaxy at $z=1$, using
an actual 2-dimensional intensity and velocity field from Fabry-Perot
observations of the local galaxy NGC 7171.  A similar project was
carried out, also using Fabry-Perot velocity fields as input, by Rix
\etal\ (1997).  NGC 7171 has a fairly typical disk velocity field
with mild spiral streaming motions, and abundant \ha\ emission, 
yielding a 2-d velocity field that samples the disk well with
high resolution.  It was observed in the \ha\ line with the
Rutgers Fabry-Perot at the CTIO 4-m Blanco telescope, on October 27
and 28, 1989 by T.B. Williams and R.A. Schommer\footnote{Cerro Tololo
Inter-American Observatory is operated by AURA under contract to the
NSF.}.  The data were reduced into intensity, velocity, and
dispersion fields by BJW using procedures described elsewhere (Palunas
and Williams 2000; Weiner \etal\ 2001) and have been previously studied
by Palunas (1996).  NGC 7171 has a heliocentric velocity
of 2740 \kms\ and we assumed a distance of 34 Mpc; its magnitude 
is $M_B=-20.0$.  The resolution of these 2-d maps is 1.5\arcsec, 
the inclination of NGC 7171 is $55\mydeg$, and its rotation velocity 
corrected for inclination is $V_c=189$ \kms.  It has a flat 
rotation curve and lies on the Tully-Fisher relation of 
Sakai \etal\ (2000).


\begin{figure*}
\begin{center}
\includegraphics[width=5.0truein]{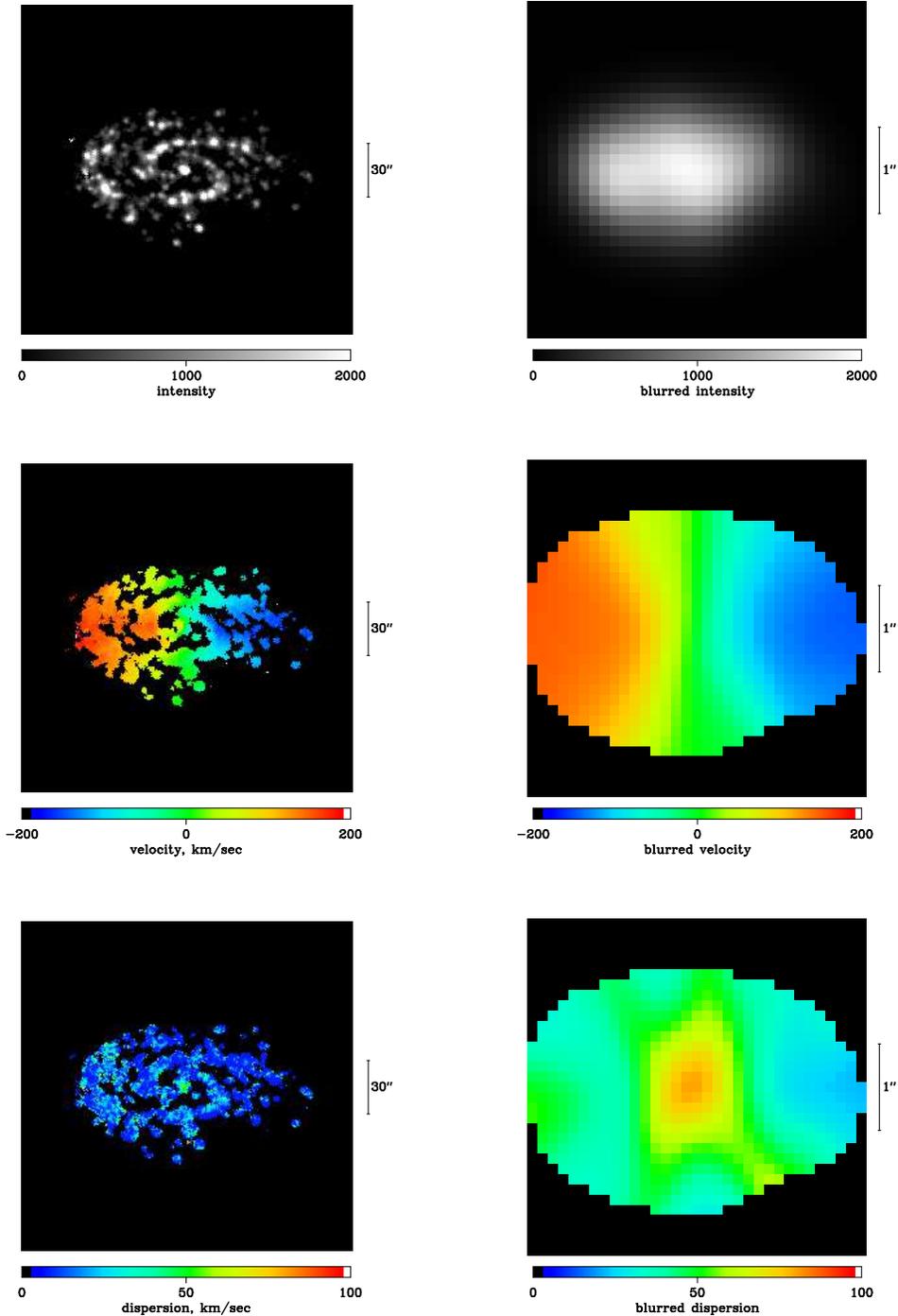}
\caption{Emission intensity, line-of-sight velocity, and dispersion 
fields for NGC 7171 as observed locally (left column) and as it would
be seen at $z=1$ with 0.7\arcsec\ seeing and
0.118\arcsec\ DEIMOS pixels (right column).  The scale bar is
30\arcsec\ on the left and 1\arcsec\ on the right.  The velocity
color scale is $\pm 200$ \kms and the dispersion color scale is
0-100 \kms.
The seeing blur smooths out the rotation velocity gradient,
but causes a central spike of dispersion where emission from
gas at different velocities is superimposed.
}
\label{fig-n7171blur}
\end{center}
\end{figure*}


The original \ha\ intensity, velocity and dispersion fields 
of NGC 7171 are shown in the left column of Figure \ref{fig-n7171blur}; 
the dispersion field is relatively featureless since
the dispersion is usually $<20$ \kms\ except in the nucleus,
where it is increased by seeing blur.  We used these fields
to reconstruct a 3-d spectral cube.  To simulate an observation
at $z=1$, we rescaled to the angular size at $z=1$, blurred each plane 
of the cube by a 0.7\arcsec\ gaussian to represent seeing, and resampled 
the cube onto DEIMOS pixels of 0.118\arcsec.  We then refit through 
the cube to produce maps of restframe intensity, velocity, and dispersion.
The right column of Figure \ref{fig-n7171blur}
shows the resulting $z=1$ maps that correspond to the
unblurred maps of the left column.

The blurred velocity and dispersion fields of Figure
\ref{fig-n7171blur} illustrate an effect we alluded to in the
discussion of rotation curve fitting in Section \ref{sec-rcfit}.  The
beam smearing due to seeing smooths out the rotation velocity
gradient, but it produces a strong peak in the velocity dispersion in
the center; where the unblurred velocity gradient is strong, the
seeing mixes gas at different velocities together.  This peak in
velocity dispersion is at high \ha\ intensity and so carries
a high weight in integrated measurements.

We used blurred
velocity cubes to simulate observations of integrated velocity
dispersion over a range of galaxy inclinations and slit-galaxy relative
position angle misalignments.  To simulate a range of inclinations, we
stretched the input NGC 7171 fields and rescaled the velocities
before blurring and resampling.  We did not allow the dispersion to
fall below 8 \kms, a typical sound speed for \hii\ regions and 
gas in the ISM.  
We did not add noise or scale the flux to a real galaxy, as
this is not a full simulation of real observations, but a test
of the relative contribution of 
emission at different velocities.  Adding noise reduces the 
detectability of spatially resolved emission at low intensities
but does not greatly affect the velocity distribution of 
the integrated emission.

We then laid down typical 1\arcsec\ wide slits over a range of relative
PAs and extracted a spectrum in a 2.3\arcsec\ (1.5 FWHM) 
extraction window, using
the intensities, velocities and dispersions of pixels within the slit
and window to build up an emission spectrum.  The resulting velocity
profile can be somewhat non-gaussian and flat-topped, especially
when the slit is aligned with the galaxy major axis, but once the
profile is convolved with the spectrograph resolution, the non-gaussianity 
is small.  We computed moments to determine the velocity dispersion 
corresponding to \sigoned.
Changing the extraction diameter within reasonable limits
had little effect on the results, because the bulk
of the signal comes from the central emission peak.  



\begin{figure}[ht]
\begin{center}
\includegraphics[width=3.5truein]{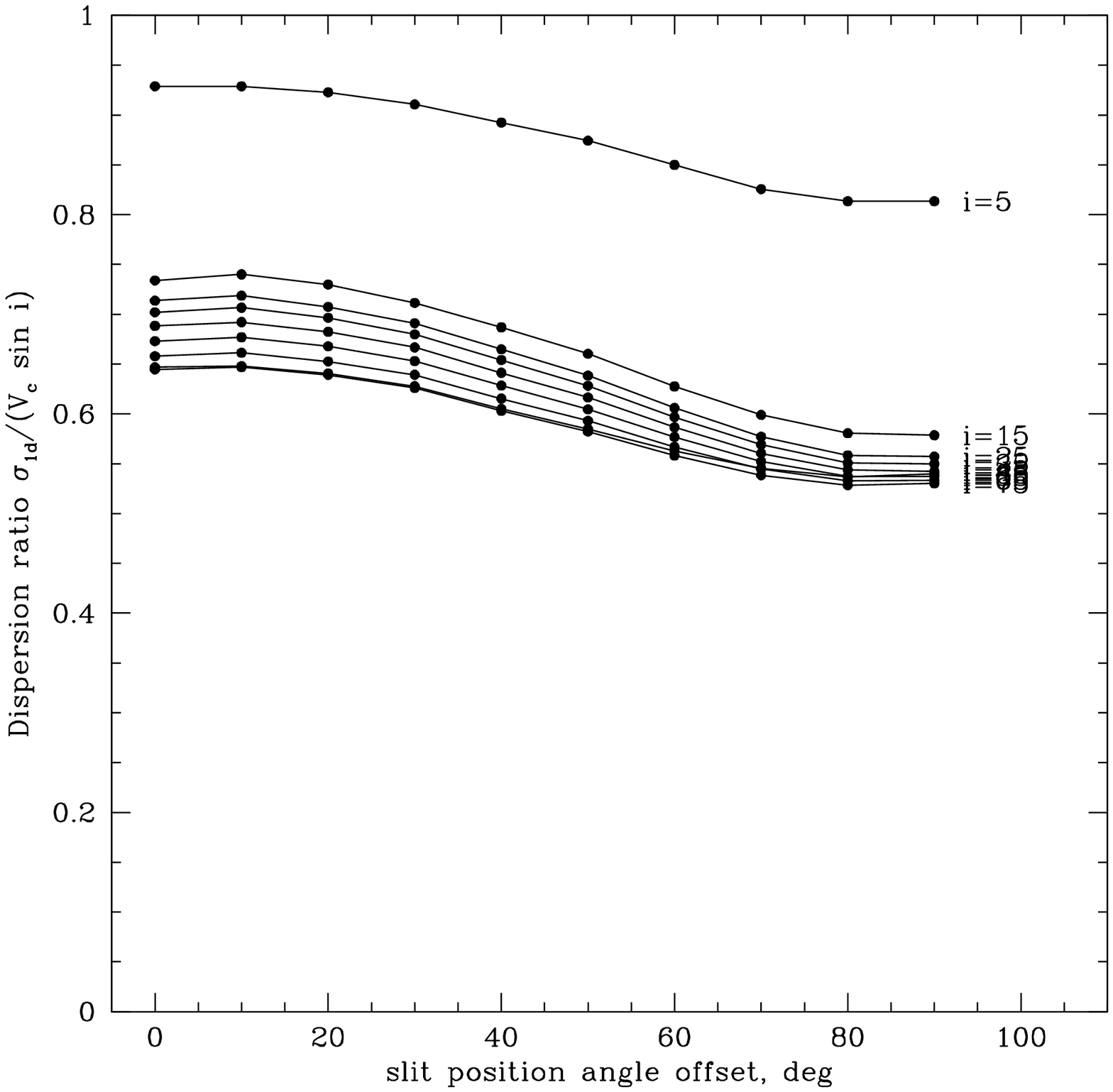}
\caption{Ratio of integrated linewidth to projected circular velocity,
$\sigoned / (V_c\, \sini)$, as a function of inclination and
slit position angle for simulated observations of NGC 7171 at
$z=1$.  At a given inclination, 
the linewidth for a perfectly misaligned slit ($\Delta PA=90$) is
only 20\% lower than the linewidth for a perfectly aligned
slit ($\Delta PA=0$).  The ratio takes out the inclination dependence,
showing that the effect of inclination other than through
\sini\ is small.  The exception is the $i=5\mydeg$ case,
where the sound speed of the gas, 8 \kms, contributes to the
observed linewidth.  }
\label{fig-simpadispratio}
\end{center}
\end{figure}



Figure \ref{fig-simpadispratio} plots the ratio of integrated
linewidth to projected circular velocity, $\sigoned / (V_c\, \sini)$,
as a function of inclination and relative slit position angle.
This ratio removes the \sini\ effect, isolating the
properties of the integrated linewidth.  There is a small residual
effect of inclination, but the variation among the
tracks is quite small.  The exception is the nearly face-on
$i=5\mydeg$ case, where the 8 \kms\ minimum dispersion contributes.
The chances of observing a face-on, perfectly flat thin disk
are very small, especially in the high-$z$ universe where many
galaxies are peculiar or irregular.  Apart from this special case,
the ratio varies by $\pm 15\%$ over the range of PA and inclination.

At a given inclination, the linewidth
for misaligned slits is lower, but the linewidth for a perfectly 
misaligned slit ($\Delta PA=90$) is only 20\% lower than the linewidth
for a perfectly aligned slit ($\Delta PA=0$).  The reason is that the 
observed linewidth comes largely from the central peak in intensity 
and dispersion, seen in the lower panel of Figure \ref{fig-n7171blur}.
The slit position angle does have a large effect on the 
observed rotation gradient, of course.  Both these effects are
seen empirically in our data in Section \ref{sec-slitpa}.
The ratio $\sigoned / (V_c\, \sini) \sim 0.6-0.65$ on average.  When
the distribution of inclinations is accounted for, the ratio is
fairly similar to the distribution in Figure 7
of Rix \etal\ (1997): $\sigoned/V_c \sim 0.55$ in the mean, with
a non-gaussian scatter due to the tail of nearly face-on model 
galaxies with low \sigoned.

The upshot of these simulations is that the seeing is a 
powerful integrator, smoothing over the disk and producing
dispersion at the location of velocity gradients, so that specific 
observational parameters like slit PA, width, and extraction 
diameter do not have a strong effect on the integrated 
linewidth.  However, the slit PA does have a strong effect on the
observed spatially-resolved rotation curve.  Similar 
conclusions were reached by Erb \etal (2004).
Inclination produces significant scatter in linewidth/velocity
ratio merely through the \sini\ factor.

A limitation is that we have modeled only a thin disk galaxy
in orderly and planar rotation.  Our comparison of 
observed rotation velocity and dispersion in Section \ref{sec-rotdispdom}
below shows that many galaxies are not dominated by orderly rotation.
However, even if the kinematics are not orderly, the effect
of the seeing as an integrator over the galaxy's emission still
operates.  The net result is that the integrated linewidth
is actually more robust than the rotation gradient against
both unusual internal kinematics and the observational effect
of slit PA offsets.

If the star formation and emission is more centrally 
concentrated than in NGC 7171, the relative contribution of high
velocity gas on the flat part of the rotation curve is reduced.
However, because the central regions do have a velocity
gradient, the integrated linewidth remains fairly high, and this
is a second order effect.  As a test, we artificially
boosted the emission intensity inside 1 kpc radius --
the central clump of emission -- by a factor of 4 and reran the 
simulations.  The integrated linewidth decreased by just 
$4.5 \pm 1\%$, over the range of inclination and slit PA.
Centrally concentrated emission must overwhelm the rest of
the galaxy to lower the linewidth significantly.

\subsection{The effect of inclination corrections and the lack thereof}
\label{sec-inclcorr}

The modeling of a blurred NGC 7171 showed that the strongest effect on
the observed integrated linewidth is the inclination factor of \sini.
Classic Tully-Fisher studies correlate velocity and luminosity after
correcting for inclination and extinction, and are generally
restricted to relatively high-inclination galaxies with clear disk
geometry.  In this study we do not restrict the sample.  In large
samples of high-redshift galaxies, correcting for inclination is only
practical with HST imaging, and galaxies may have stronger deviations
from an ideal thin circular disk geometry, so the relation between 
ellipticity and inclination is less direct.  As groundwork for the 
Tully-Fisher study of Paper II, here we model the effect on disk
Tully-Fisher relations when inclination and extinction corrections 
are omitted.  


\begin{figure}[ht]
\begin{center}
\includegraphics[width=3.5truein]{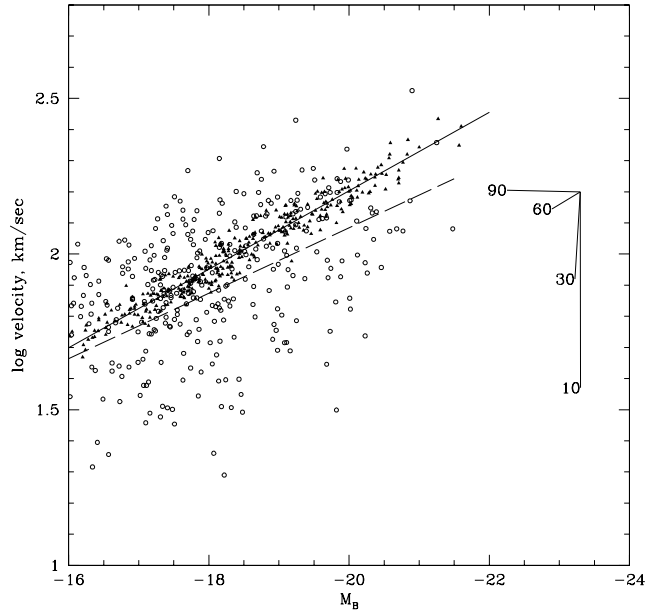}
\caption{A simulated Tully-Fisher relation illustrating the 
effect of omitting inclination and extinction corrections.
The black triangles are drawn from a Tully-Fisher relation 
with inverse TF slope -0.125 dex/mag and intrinsic scatter 0.0375 dex.
The solid diagonal line is a linear fit to these points.
The open circles are the same galaxies ``observed''  
at randomly assigned inclinations $i$.  
The decorrected quantities are velocities $V_{rot}~{\rm sin}~i$ with 
0.084 dex observational error added
and magnitudes $M_B+A_B(i)$ with 0.11 dex error.  
The dashed diagonal line is a linear fit to the decorrected points.
The lines at right indicate the decorrection tracks for a galaxy with 
log~$V_{rot}=2.2$ at $i=10\mydeg,30\mydeg,60\mydeg,90\mydeg$.  
Edge-on galaxies move above the baseline inverse-TF relation due to 
extinction; face-on galaxies fall below the baseline due to low 
projected velocities.  For $B$ band extinction, galaxies at the 
median inclination $i=60$ are nearly unchanged.  However,
the scatter around the TF relation becomes large and non-gaussian.
}
\label{fig-inclsimul}
\end{center}
\end{figure}


Figure \ref{fig-inclsimul} shows a model Tully-Fisher relation 
with and without inclination and extinction corrections.
The filled triangles are a simulated ``true'' Tully-Fisher
relation with a realistic distribution of magnitudes (drawn
from our sample in $0<z<0.5$), an inverse-TF log~$V(M)$ slope of 
$-0.125$ dex/mag, and an intrinsic scatter of 0.0375 dex (0.3 mag).
Each galaxy is assigned an inclination 
drawn from a uniform distribution in ${\rm cos}~i$; we do not
modify the inclination distribution to reflect
the small bias against edge-on galaxies in a magnitude-selected
sample, and surface brightness selection against face-on 
galaxies.  The open circles are the decorrected values log~$V_{rot,obs}$
and $M_{B,obs}$, with errors added from gaussian distributions
of 0.084 dex and 0.11 mag.  The extinction correction (to face-on) is 
$A_B=\gamma_B ~{\rm log}(a/b)$, 
with $\gamma_B=1.57$ (Sakai \etal\ 2000, neglecting the 
luminosity dependence of extinction), and we assume the galaxies have
intrinsic thickness:radius 1:5.  The velocity is $V_{rot}~{\rm sin}~i$
with a random-motion component of 25 \kms\ added in quadrature, 
which prevents the face-on galaxies' velocity from falling to near
zero.  The solid and dashed lines show the best linear fits
respectively to the original and decorrected points with $M_B<-18$
(using a fit method that compensates for scatter, described
in Paper II).  The fitted slopes are $-0.126\pm 0.003$ for the 
original sample and $-0.105 \pm 0.021$ for the decorrected sample.

The lines at right show how decorrection moves a galaxy with
log~$V_{rot}=2.2$ for inclinations $10\mydeg,30\mydeg,60\mydeg,90\mydeg$.
The undoing of these inclination and extinction corrections
moves the galaxies onto tracks nearly parallel to the original
TF relation -- e.g. all the $i=90\mydeg$ galaxies are shifted to fainter
magnitudes, so appear above the original TF.   For $\gamma_B=1.57$
and a TF slope $-0.125$, the corrections cancel at $i=60.5\mydeg$,
which fortuitously is almost exactly the median inclination 
of $i=60\mydeg$ for a randomly oriented sample. 
Thus in $B$-band, the effect of undoing corrections on the TF intercept
is relatively small.  For the galaxies on a single inclination track,
the slope is unchanged, but for the entire sample,
the slope of the uncorrected inverse TF relation is shallower.

As shown in Figure \ref{fig-inclsimul}, the scatter induced
by lack of corrections is non-gaussian, although the non-gaussianity 
is blurred somewhat by the observational errors.  The effect
on Tully-Fisher fitting depends on the magnitude distribution
and limit of the sample.  The fitted TF relation changes due to 
the correlated anti-corrections scattering across the TF ridgeline, the
non-gaussian scatter, and the elimination of edge-ons by the magnitude 
limit.  For the realization shown here,
a fit to the galaxies brighter than $M_{B,obs}=-18$ yields a 
velocity intercept shifted down by 0.11 dex at $M_{B,obs}=-20$, 
a slope of $-0.105 \pm 0.021$,  and a greatly increased 
scatter of 0.19 dex.  We emphasize that 
these values only apply for the idealized thin disk assumptions, which
likely do not apply to many galaxies in the TKRS -- for example,
elongated but non-rotating galaxies like TKRS 5627 and 10023,
shown in Figures \ref{fig-exampleimages} and \ref{fig-examplespec}.  

Local Tully-Fisher samples require a luminosity or velocity-dependent
extinction correction; extinction is higher in larger galaxies
(Giovanelli \etal\ 1995; Tully \etal\ 1998).  Undoing this
correction means that an uncorrected inverse TF sample has
a $V(M)$ slope steeper than the true slope.  In the $B$ band
the effect is about 10-15\% of the magnitude range of the sample
(Tully \etal\ 1998), so the uncorrected inverse TF slope is $\sim 10-15\%$
higher.  This could counteract the slope-shallowing effect described
above.  However, one must tread carefully when applying local
extinction corrections to distant samples.  Both the extinction effect
and the inclination-induced offsets are reasons to attempt 
comparisons internal to the sample, spanning a range of redshifts.

The general effects of the lack of correction are
small shifts in intercept and slope, and a substantial
increase in scatter.  
If the TKRS galaxies with linewidths were all close to
the thin rotating disk model,
applying inclination and extinction corrections would greatly
reduce the scatter in the Tully-Fisher relations fitted in Paper II.
In fact, in Paper II we find that applying either the 
\sini\ inclination correction, or both inclination and
extinction corrections, {\it does not} reduce the scatter
in Tully-Fisher relations.
A small amount of this can be due to statistical errors on 
inclination, but most is due to the diversity of kinematic
properties, galaxies with non-disk shapes, and
motions that are not strictly rotation, as we show 
in Section \ref{sec-spatialkin}.

\subsection{Empirical dependence on slit position angle}
\label{sec-slitpa}

Rotation curves of inclined disks should be harder to measure 
when the slit is not aligned close to the major axis.  The
modeling of NGC 7171 suggests that integrated velocity linewidths
are not as severely affected by slit misalignments.
The effect can be tested empirically for the TKRS sample since
HST imaging allows measures of the galaxy position angles.

We used a compilation of measurements from ellipse fitting
to the ACS $i$-band galaxy images, yielding ellipticity,
position angle, and half-light radius.  Segmentation
images output by the SExtractor program (Bertin \& Arnouts 1996)
were used to mask out neighboring galaxies and the IRAF task
{\tt ellipse} was run to fit elliptical 
isophotes.\footnote{IRAF is distributed by NOAO, which is 
operated by AURA, Inc., under a cooperative agreement with 
the National Science Foundation.}
Further results on the galaxy sizes will be reported separately
(Melbourne \etal\ 2006).
Here we use the ellipticities and position angles to determine
slit misalignments.


\begin{figure}[ht]
\begin{center}
\includegraphics[width=3.5truein]{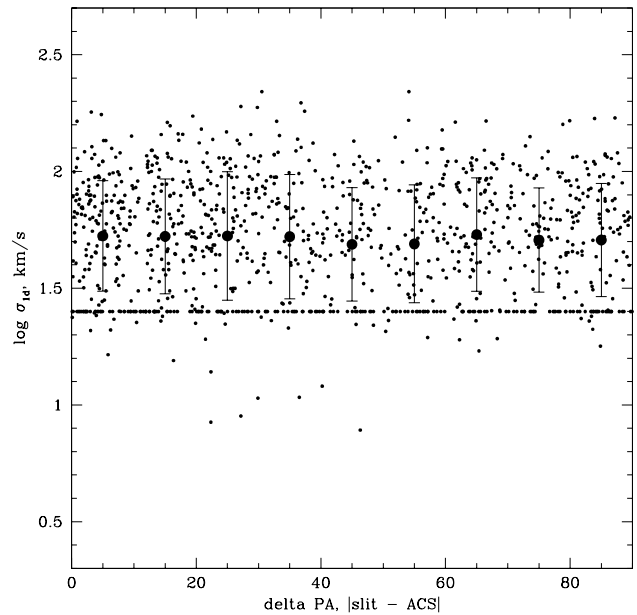}
\caption{Integrated linewidth 
\logsigoned\
as a function 
of position angle misalignment: offset between slit PA and 
major axis PA from ellipse fits to HST/ACS images.  
Large points show the mean and RMS in bins of angle.
The small points at $\logsigoned=1.4$ are the kinematically 
unresolved objects.
There is no visible trend of measured 1-d linewidth with PA offset.
}
\label{fig-pasigma}
\end{center}
\end{figure}


Figure \ref{fig-pasigma} shows integrated linewidth \logsigoned\
as a function of the misalignment between slit and image position
angle.  The large points show the mean and RMS in bins of PA offset.
(The slitmasks were designed based on PAs from ground based imaging
before the HST catalog was available, so the slits are not exactly
aligned with HST PAs.)
Slit misalignment has no visible effect on linewidth measured
from the 1-d spectrum.  A fraction of the galaxies in this
plot are nearly round, so that the PA is not very meaningful;
removing these objects does not change the conclusion. 
A small trend with PA offset might be expected from the models
in Figure \ref{fig-simpadispratio}, but galaxies that are not
orderly rotating disks, and galaxies whose ellipticity and
kinematics are misaligned, will weaken such a trend.


\begin{figure}[ht]
\begin{center}
\includegraphics[width=3.5truein]{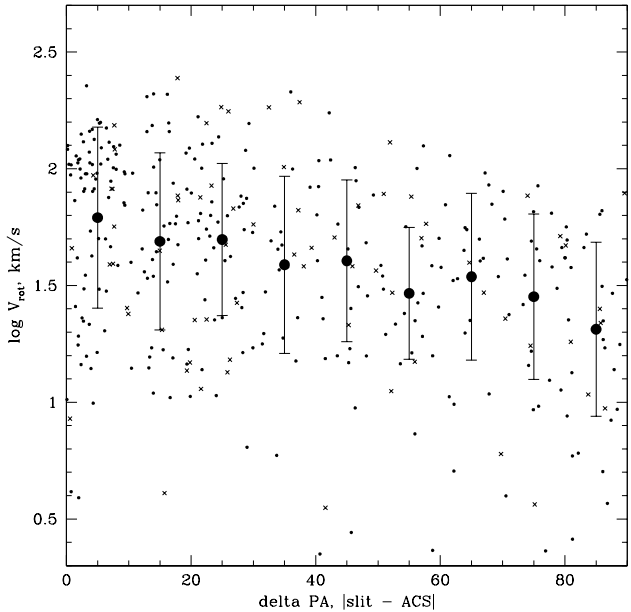}
\caption{Rotation velocity log \vrot\ as a function 
of position angle misalignment, offset between slit PA and HST/ACS major 
axis PA.  Round galaxies with ellipticity $e<0.25$ are plotted as
Xes, non-round galaxies are circles.
Large points show the mean and RMS of non-round galaxies in bins of angle.
Extracted rotation velocities are higher for aligned slits; there is
a deficit of high \vrot\ for misaligned slits.
}
\label{fig-pavrot}
\end{center}
\end{figure}


For the subsample of galaxies with 2-d rotation fits,
Figure \ref{fig-pavrot} shows the rotation velocity log \vrot\
as a function of the slit misalignment angle.  Again, large
points show the mean and RMS in bins of PA offset.  Here, there is
a strong effect: measured rotation velocity declines as misalignment 
increases, and there is a lack of high \vrot\ objects for misaligned 
slits, as expected.  However, note that even aligned slits have a fair
number of low \vrot\ objects.  

Clearly, measuring rotation velocity from a misaligned slit is
subject to large error, especially since {\sc ROTCURVE} does not use
a full 2-d model.  For purposes of studying the properties of
rotation velocity and dispersion in the next section,
we split the sample into ``aligned'' and ``misaligned'' 
with $|\Delta PA| <40\mydeg$ and $>40\mydeg$.
We also exclude the nearly-round galaxies with 
ellipticity $e<0.25$ from the ``aligned'' subsample.

\section{Properties of spatially resolved kinematics}
\label{sec-spatialkin}

\subsection{Rotation versus dispersion dominated galaxies}
\label{sec-rotdispdom}

The two galaxies whose rotation and dispersion 
profiles are shown in Figure \ref{fig-examplerc} exemplify
a trend we found in the entire sample.  In order to
fit the kinematics, we had to include both velocity \vrot\
and dispersion \sigtwod\ in the 2-d modeling of {\sc ROTCURVE},
discussed in Section \ref{sec-rotcurve}.  Quite a few galaxies
like TKRS 5627 and 10023 show a low rotation velocity but a significant 
dispersion.  

Table \ref{table-rotcatalog} lists galaxy ID,
redshift, slit PA offset, and the fitted line-of-sight 
rotation and dispersion for the 380 galaxies with 
good {\sc ROTCURVE} fits.  It also includes morphological
classifications for a subset of the galaxies, discussed further
in Section \ref{sec-rotdispnature}.  Because there are galaxies
which appear not to be consistent with inclined thin
rotating disks, we report the line-of-sight velocities
rather than applying an ellipticity-based inclination
correction.  If a significant number of high-redshift
galaxies are not circular rotating disks, the chain of reasoning
from axis ratio to inclination to \sini\ correction breaks down.

\begin{deluxetable}{rrrrrl}

\tablecaption{
Catalog of TKRS rotation and dispersion fits
\label{table-rotcatalog}
}

\tablecolumns{6}
\tablewidth{0pt}
\tabletypesize{\scriptsize }

\tablehead{
 TKRS ID & $z$\tablenotemark{1} & $|\Delta PA|$\tablenotemark{2} & 
 \vrot\tablenotemark{3} & \sigtwod\tablenotemark{4} & morphology\tablenotemark{5}
 }
 \startdata
0000428 & 0.4870 & 13.2 & 14.0 & 15.0 & hyphen \\
0000448 & 0.4724 & 70.1 & 80.0 & 40.1 & --- \\
0000555 & 0.4589 & 81.0 & 37.8 & 69.6 & --- \\
0000714 & 0.9434 & 52.8 & 21.6 & 48.8 & --- \\
0000760 & 0.5032 & 14.0 & 209.2 & 10.0 & disk \\
0001126 & 0.9447 & 25.8 & 62.9 & 64.3 & irregular \\
0001217 & 0.4587 & 0.8 & 4.1 & 59.4 & merger \\
0001226 & 0.5023 & 13.7 & 0.0 & 65.0 & irregular \\
0001289 & 0.1137 & 54.5 & 56.5 & 69.9 & --- \\
0001333 & 1.2960 & 80.1 & 46.9 & 74.4 & --- \\
0001432 & 0.7479 & 48.6 & 0.6 & 54.3 & --- \\
0001563 & 0.3758 & 16.4 & 157.0 & 5.0 & edge-on \\
0001577 & 0.4855 & 2.1 & 110.2 & 5.0 & disk \\
0001769 & 0.1192 & 66.3 & 50.0 & 65.1 & --- \\
0001808 & 0.0787 & 21.5 & 64.3 & 39.8 & edge-on \\
0001861 & 1.3630 & 3.0 & 95.1 & 40.0 & disk \\
0001923 & 0.2755 & 43.0 & 50.8 & 39.9 & --- \\
0001950 & 1.3640 & 45.3 & 21.4 & 89.5 & --- \\
0002011 & 0.4734 & 88.4 & 9.3 & 24.8 & --- \\
0002012 & 1.0010 & 0.4 & 104.4 & 55.0 & disk \\

 \enddata

\tablecomments{The complete version of this table is in the electronic edition 
of the Journal.  The printed edition contains only a sample.}

\tablenotetext{1}{Redshift.}
\tablenotetext{2}{Absolute value of the offset between DEIMOS slit position 
angle and photometric major axis PA in HST/ACS $I$ images, degrees.}
\tablenotetext{3}{\,Line-of-sight rotation velocity from seeing-compensated 
modeling of the 2-d spectra, \kms.  For $|\Delta PA| \gtrsim 40\mydeg$,
rotation velocities are strongly affected by slit misalignment.}
\tablenotetext{4}{\,Line-of-sight dispersion from seeing-compensated modeling
of the 2-d spectra, \kms.  Values are near-quantized in units of 5 \kms\
due to the model grid.}
\tablenotetext{5}{Morphological type from visual classification.  Only
galaxies with $|\Delta PA|<40$ and ellipticity $>0.25$ were classified.}

\end{deluxetable}


The simulated seeing-blurred velocity fields in Section 
\ref{sec-blurdisk} suggest that observing a rotating galaxy
with the slit along the minor axis yields a low velocity
gradient but a central peak in velocity dispersion.  These
are found in our sample, but TKRS 5627 and 10023, shown
in Figures \ref{fig-exampleimages}, \ref{fig-examplespec},
and \ref{fig-examplerc}, are different.  They exemplify a class
of galaxies which have low rotation, high dispersion, and
aligned slits ($|\Delta PA|<40\mydeg$), so that the lack of
rotation is not due to slit misalignment.  These galaxies
also frequently show dispersion profiles which are not sharply
peaked, although the limited spectral resolution of TKRS 
makes dispersion profiles rather noisy.  

\begin{figure}[ht]
\begin{center}
\includegraphics[width=3.5truein]{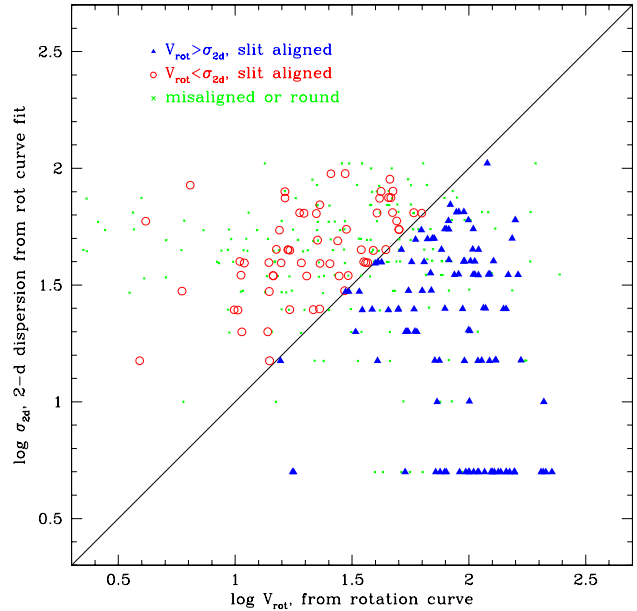}
\caption{2-d dispersion versus rotation velocity for galaxies with
2-d ROTCURVE modeling.
Open circles: dispersion dominated galaxies, $\vrot<\sigtwod$, with $e>0.25$.
Filled triangles: rotation dominated galaxies, $\vrot>\sigtwod$, with $e>0.25$.
Small Xes: round galaxies or misaligned slits.
We divide the galaxies into rotation and dispersion dominated 
at $\vrot=\sigtwod$.  The values of \sigtwod\ are quantized in steps
of 5 \kms\ by the fitting method.}
\label{fig-vrotvdisp}
\end{center}
\end{figure}

We classified galaxies into rotation- and dispersion-dominated,
based on whether $\vrot > \sigtwod$ or vice versa.  Figure 
\ref{fig-vrotvdisp} plots \sigtwod\ versus \vrot; the placement
of the dividing line is somewhat arbitrary, but intended to
separate galaxies whose main kinematic support is from orderly
rotation, and galaxies where dispersion (or the disordered gas
motions that \sigtwod\ may represent) plays a major role.
When roundish
galaxies with $e<0.25$ and those with misaligned slits,
$|\Delta PA|>40\mydeg$, are excluded, there are 185 galaxies
with good 2-d fits from {\sc ROTCURVE}.  Of these, 118 are
rotation dominated and 67 (36\%) are dispersion dominated.
Thus two-thirds of the subsample have kinematics basically 
similar to local rotating galaxies -- which was by no means
guaranteed -- but
over one-third of this subsample have kinematics that are
inconsistent with a thin rotating disk model.

The nature of the dispersion dominated galaxies is a puzzle.  They are
not literally dispersion-supported elliptical galaxies.  
Comparing rotation and dispersion dominated galaxies
(RDGs and DDGs) in magnitude and color, the DDGs are only slightly
different, 0.05 bluer in median $U-B$ and equal in median brightness,
and their morphologies are late type; we discuss DDG and RDG
properties further in Section \ref{sec-rotdispnature}.  The DDGs'
velocity dispersion is unlikely to indicate a literal dynamically hot 
pressure-supported tracer population like stars in spheroidals.
We are measuring velocities from nebular lines that come from 
$\sim 10^4$ K gas, so the dispersion in individual \hii\ regions
is only 8-10 \kms.  
The larger values of \sigtwod\ found in DDGs could come from the
relative motions of the ensemble of gas clouds -- chaotic,
disordered, or out-of-plane velocity fields that are smoothed over 
by the seeing, or that have strong misalignments between photometry 
and kinematics.  However, the geometry of the motions and whether 
they will eventually dissipate and settle into a disk cannot be
resolved in these spectra.

If the seeing blur caused a bias against the detection of 
rotation in small objects, it could make objects appear
dispersion dominated by masking the true rotation gradient.
In principle, the {\sc ROTCURVE} fitting method is 
seeing-compensated.  For a very small object with a 
rotation gradient that suffers greatly from seeing blur, 
it yields a shallow minimum in $\chi^2$; models with an intrinsic
rotation gradient and with a dispersion both fit the data, so
there is a large error on \vrot\ rather than a bias.
In practice, we rejected objects with a very small diameter
of velocity data, $\lesssim 1\arcsec$.  In Section 
\ref{sec-rotdispnature} we discuss the sizes of RDGs and
DDGs and show that the DDGs are only mildly smaller, so
they are not simply caused by an observational bias against
rotation.  Intrinsic non-roundness or errors in the
ellipticity and PA measurements could cause a few face-on
disks to appear in the DDG sample, but there are too many
elongated DDGs to explain away.


\begin{figure}[ht]
\begin{center}
\includegraphics[width=3.5truein]{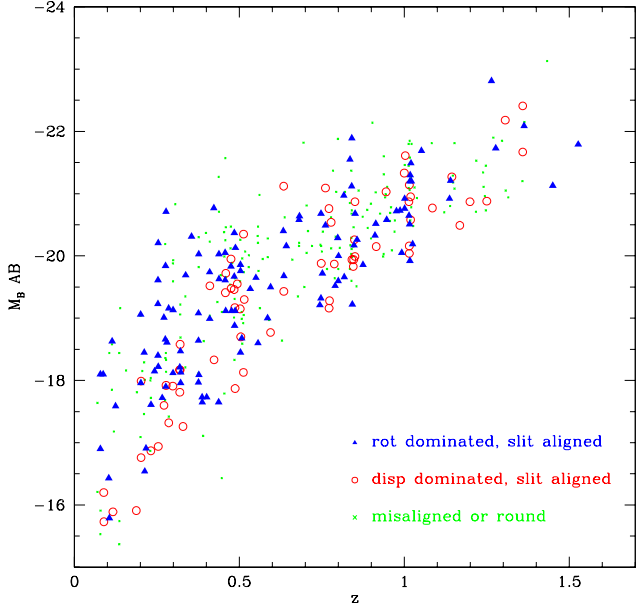}
\caption{Magnitude versus redshift for galaxies with
2-d ROTCURVE modeling.
Open circles: dispersion dominated galaxies, $\vrot<\sigtwod$, with $e>0.25$.
Filled triangles: rotation dominated galaxies, $\vrot>\sigtwod$, with $e>0.25$.
Small Xes: round galaxies or misaligned slits.
Dispersion dominated galaxies exist at low redshift, but are
relatively faint.  There is not strong evidence
for redshift evolution in the rotation/dispersion fraction, but 
possibly bright dispersion dominated galaxies are only found 
at high redshift.}
\label{fig-zmagrotdispdom}
\end{center}
\end{figure}


Dispersion dominated galaxies occur over the full
redshift range of our sample
and are not strictly a high-redshift population.
Figure \ref{fig-zmagrotdispdom} plots magnitude against
redshift for the sample with 2-d {\sc ROTCURVE} modeling,
 with points coded for rotation or dispersion
domination.  DDGs exist at low redshift, but are
relatively faint there; it is possible that the dispersion/rotation 
fraction is higher at high redshift, although the sample size 
and selection effects make any conclusion preliminary.

\subsection{Comparison of integrated linewidths and spatially resolved kinematics}
\label{sec-rotdispsigmacomp}

A key question for using linewidths of integrated emission as a
characteristic velocity scale is how well they correlate with
spatially resolved kinematic measures.  In Section \ref{sec-blurdisk}
we tested models of seeing-blurred integrated emission from a 
rotating disk and found that linewidth \sigoned\ correlated well
with projected circular velocity $V_c \sini$; the chief scatter 
between \sigoned\ and true $V_c$ is induced by inclination.  
In this section we test the relation of linewidth to 
rotation and dispersion from spatially resolved fits empirically using
the 1-d and 2-d kinematic fits in TKRS.  These quantities are
line-of-sight, so no inclination correction is applied.


\begin{figure}[ht]
\begin{center}
\includegraphics[width=3.5truein]{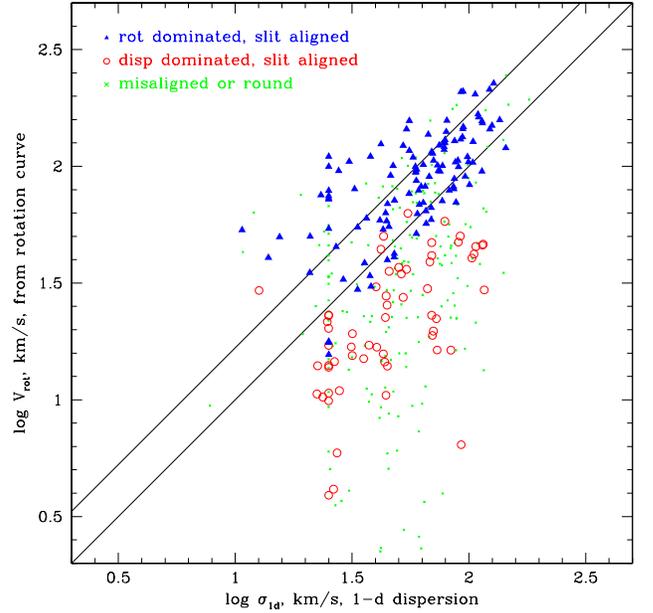}
\caption{Rotation velocity from the 2-d ROTCURVE modeling,
log \vrot, versus integrated linewidth \logsigoned.
Open circles: dispersion dominated galaxies, $\vrot<\sigtwod$, with $e>0.25$.
Filled triangles: rotation dominated galaxies, $\vrot>\sigtwod$, with $e>0.25$.
Small Xes: round galaxies or misaligned slits.
The galaxies at $\logsigoned=1.4$ are kinematically unresolved
in the 1-d linewidth.
The diagonal lines are the 1:1 line and the Rix \etal\ (1997)
$\sigma = 0.6 V_c$ line.
For dispersion dominated galaxies, \sigoned\ captures the velocity
scale but \vrot\ goes to small values.}
\label{fig-compsigvrot}
\end{center}
\end{figure}

\begin{figure}[ht]
\begin{center}
\includegraphics[width=3.5truein]{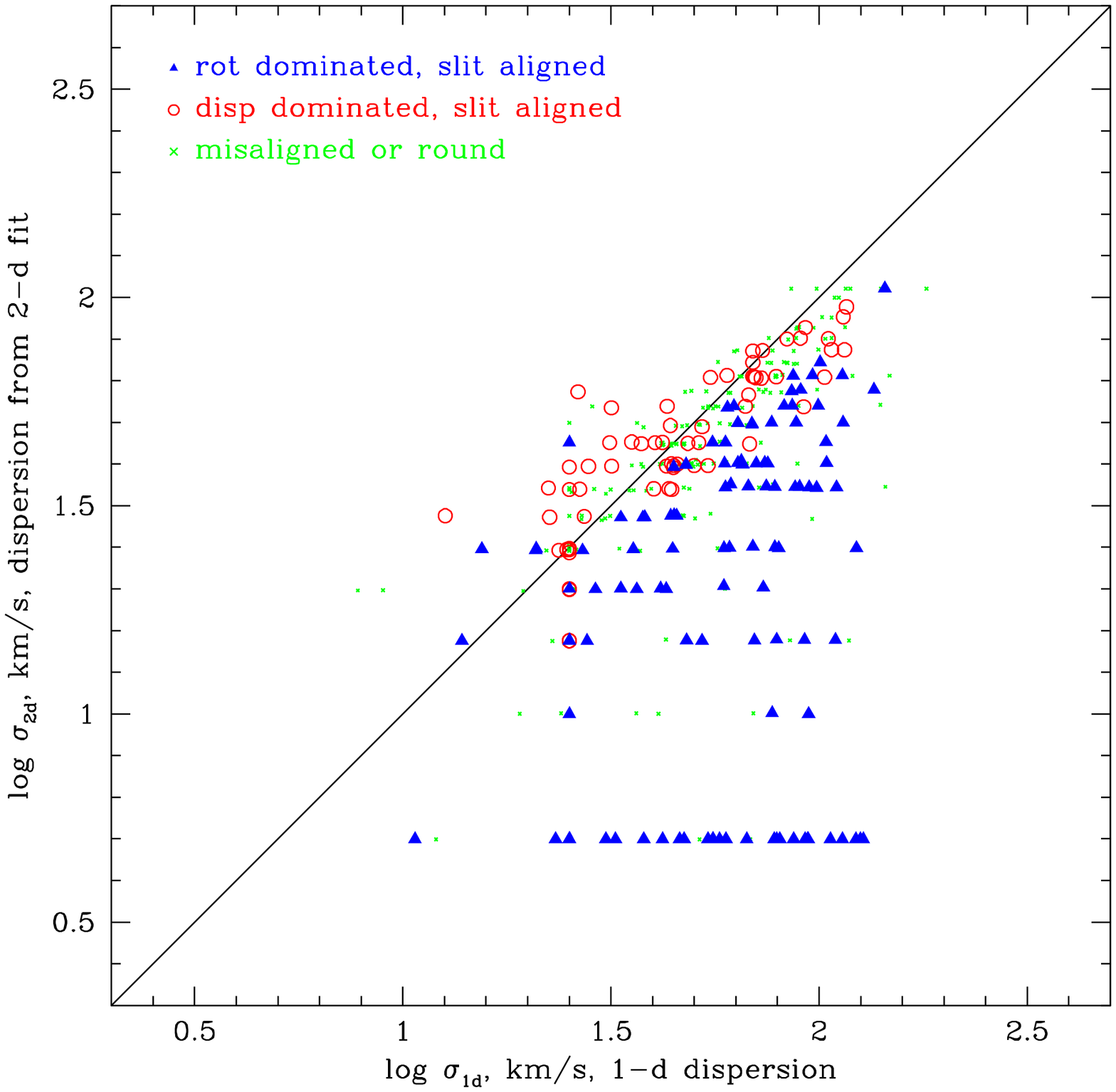}
\caption{Dispersion from the 2-d ROTCURVE modeling, log \sigtwod,
versus integrated linewidth \logsigoned.  The
\sigtwod\ values are quantized by fitting on a 5 \kms\ model grid.
Open circles: dispersion dominated galaxies, $\vrot<\sigtwod$, with $e>0.25$.
Filled triangles: rotation dominated galaxies, $\vrot>\sigtwod$, with $e>0.25$.
Small Xes: round galaxies or misaligned slits.
The diagonal line is the 1:1 line.
For dispersion dominated galaxies, \sigoned\ and \sigtwod\ are
correlated; for rotation dominated galaxies, \sigoned\ captures
the velocity scale while \sigtwod\ becomes small.}
\label{fig-compsigvdisp}
\end{center}
\end{figure}


The existence of rotation and dispersion dominated galaxies
suggests that neither 2-d measure, \vrot\ nor \sigtwod, by
itself adequately represents the velocity scale of every distant
galaxy.  Figures \ref{fig-compsigvrot} and \ref{fig-compsigvdisp}
plot the 2-d velocity and dispersion, \vrot\ and \sigtwod,
against the 1-d integrated linewidth, \sigoned.  Point types
are coded by whether the galaxy's 2-d kinematics are rotation or
dispersion dominated.  The clustering of points with 
$\logsigoned=1.4$ are galaxies kinematically unresolved in 1-d.

Figures \ref{fig-compsigvrot} and \ref{fig-compsigvdisp} are
the flip-side of each other.  Each 2-d quantity, when it
dominates, correlates with the 1-d measure \sigoned, but 
falls well below \sigoned\ when it does not dominate.  Either
used alone would leave some galaxies with egregiously small velocities.

For rotation dominated galaxies, there is a fair amount of scatter
between \vrot\ and \sigoned, but in general these quantities are 
correlated, with \sigoned\ offset below \vrot.  But for dispersion
dominated galaxies, \vrot\ is low and does not reflect the 
kinematic support seen in \sigoned.
Conversely, \sigtwod\ correlates well with \sigoned\ in dispersion
dominated galaxies.  But rotation dominated galaxies can be fit
without requiring 2-d dispersion, so \sigtwod\ falls below
\sigoned.  In the RDGs, \sigoned\ is capturing the seeing-induced
dispersion.


\begin{figure}[ht]
\begin{center}
\includegraphics[width=3.5truein]{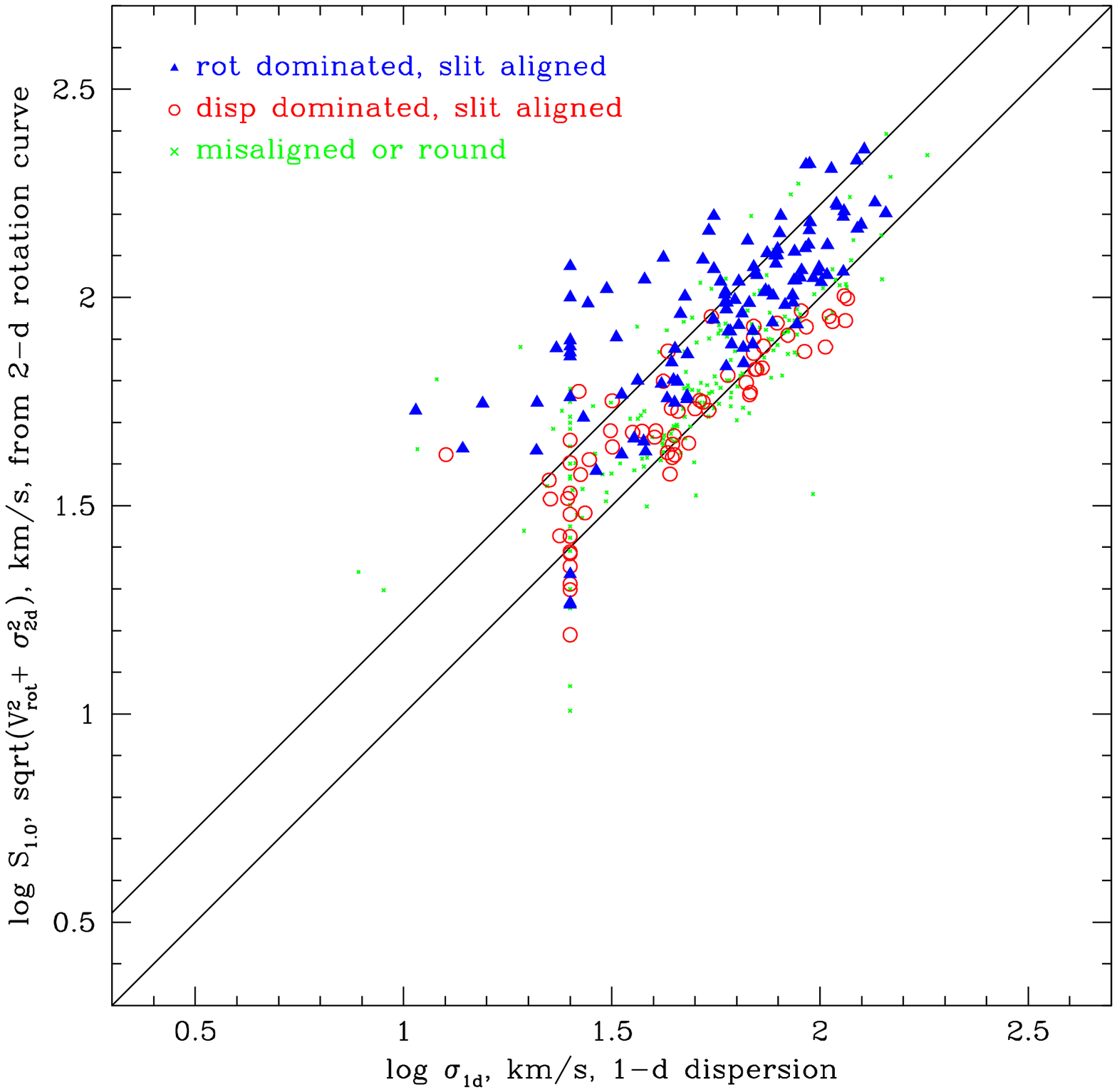}
\caption{Combined velocity, $\wone^2 = \vrot^2 + \sigtwod^2$:
log \wone\ versus integrated linewidth \logsigoned.
Open circles: dispersion dominated galaxies, $\vrot<\sigtwod$, with $e>0.25$.
Filled triangles: rotation dominated galaxies, $\vrot>\sigtwod$, with $e>0.25$.
Small Xes: round galaxies and those 
with large error on \vrot\ and \sigtwod.
The diagonal lines are the 1:1 line and the Rix \etal\ (1997)
$\sigma = 0.6 V_c$ line.  Linewidth and \wone\ are correlated,
falling closer to the 1:1 line for dispersion dominated and to the 
Rix \etal\ line for rotation dominated galaxies.}
\label{fig-compsigvcomb}
\end{center}
\end{figure}

\begin{figure}[ht]
\begin{center}
\includegraphics[width=3.5truein]{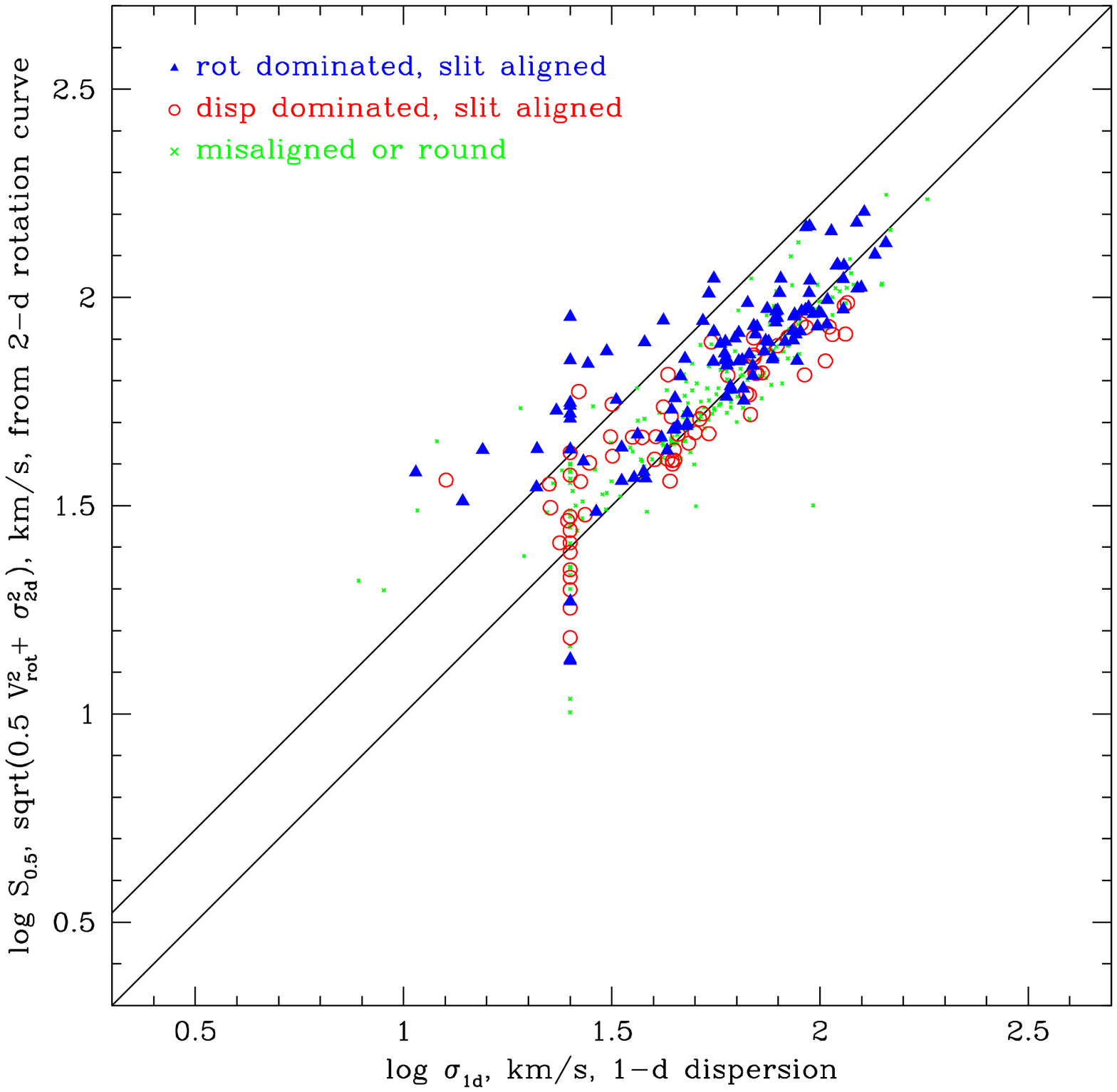}
\caption{Combined velocity, $\whalf^2 = 0.5 \vrot^2 + \sigtwod^2$:
log \whalf\ versus integrated linewidth \logsigoned.
Open circles: dispersion dominated galaxies, $\vrot<\sigtwod$, with $e>0.25$.
Filled triangles: rotation dominated galaxies, $\vrot>\sigtwod$, with $e>0.25$.
Small Xes: round galaxies and those 
with large error on \vrot\ and \sigtwod.
The diagonal lines are the 1:1 line and the Rix \etal\ (1997)
$\sigma = 0.6 V_c$ line.  Linewidth and \whalf\ are correlated;
the 0.5 prefactor makes the combined velocity width a better estimate of
velocity dispersion, so that the correlation is tighter and the galaxies 
closer to the 1:1 line.}
\label{fig-compsigvcombtwo}
\end{center}
\end{figure}


The inadequacy of each 2-d quantity \vrot\ and \sigtwod\ by
itself motivated the design of {\sc ROTCURVE}, which fits
both simultaneously.  To represent a characteristic velocity
derived from the 2-d {\sc ROTCURVE} fits, we define a 
``combined velocity scale'' $S_K$:

\begin{equation}
S_K^2 = K \vrot^2 + \sigtwod^2,
\end{equation}

\noindent
where K is a constant $\leq 1$.  The scale $S_K$ combines
rotation and pressure or random/disordered-motion support, just as
Tully-Fisher studies have done to model an observed
integrated \hi\ linewidth (e.g. Tully
\& Fouque 1985).  The simplest combination is \wone, with 
$K=1$.  This is in some sense an estimator of rotation 
velocity plus a turbulent-motion correction.

However, setting $K<1$ makes $S_K$ a better estimator
of the velocity dispersion $\sigma$ of the entire potential,
because rotation velocity and dispersion contribute differently
to the support of a gravitationally bound system.
For example, in an isothermal potential with a flat rotation
curve, a population with isotropic velocity dispersion and
no net rotation has
$\sigma = V_c/\sqrt{2}$, suggesting that with $K=0.5$,
\whalf\ is a good estimator of $\sigma$ of the potential.
In general, for a spherically symmetric tracer distribution with 
density $\propto r^{-\alpha}$ and isotropic velocity dispersion, 
if $V_c$ varies slowly with radius, $\sigma = V_c/\sqrt{\alpha}$ 
(Binney \& Tremaine 1987).  
Real galaxies in the range at or beyond the peak of the
baryonic contribution to the rotation curve,
where rotation curves are measured, have $\alpha = 2 - 3$ for
a spherized mass distribution, so $K=0.3-0.5$ is reasonable.

Figures \ref{fig-compsigvcomb} and \ref{fig-compsigvcombtwo}
plot the combined velocity scales \wone\ and \whalf\ against
1-d linewidth \sigoned.  Combined velocity alleviates 
the shortcomings of \vrot\ and \sigtwod, eliminating artificially
low velocity measurements.  Both \wone\ and \whalf\ show
strong correlations with \sigoned.  For the subsample with aligned slits
and $e>0.25$, the mean offset ${\rm log}~\wone - \logsigoned$
is 0.15 dex with RMS 0.17 dex; for \whalf\ the mean and RMS log offset are
0.06 and 0.14.  The rotation estimator \wone\ shows its
rotation velocity heritage; rotation dominated galaxies have 
$\wone>\sigoned$, relatively close to the $\sigma = 0.6 V_c$
line of Rix \etal\ (1997).  RDGs and DDGs are offset in \wone;
the offset in \whalf\ is smaller.
When the dispersion-analog \whalf\ is plotted against \sigoned,
the correlation with \sigoned\ tightens, and the offset between
\whalf\ and \sigoned\ is reduced, because both are estimating
a velocity dispersion of the potential.

The good correlation between 1-d integrated linewidth and 
the combined velocity scale from 2-d rotation and dispersion
fits has two major implications.  First, 1-d linewidth is a 
reliable representation of the more sophisticated 2-d kinematics.
As suggested by the simulations in Section \ref{sec-blurdisk}, the
seeing is a powerful integrator.  The 1-d linewidth can be used
as a kinematic measure when estimating dynamical masses, although
the $\sigoned \sim 0.6V_c$ and the substantial scatter should
be kept in mind.  Second, 2-d rotation 
velocity alone does not completely account for the kinematic
support for all galaxies.  For galaxies that are selected 
to have visibly rotation-dominated kinematics, \vrot\ is
adequate, but for broader samples, \vrot\ alone will not
suffice.  Constructing a Tully-Fisher
relation using \vrot\ only will push galaxies with
$\sigtwod>\vrot$ to erroneously low velocity, as seen in
Figure \ref{fig-compsigvrot}.  

\section{Discussion: The Nature of Rotation and Dispersion Dominated Galaxies}
\label{sec-rotdispnature}

The existence of rotation and dispersion dominated emission line
galaxies -- essentially two modes of dynamical support, or a
sort of kinematic morphology --  raises the question of their
nature and evolution.  Rotation is well understood in local
galaxies; dispersion dominated galaxies are the real puzzle.  
The size of the TKRS sample yields enough RDGs and DDGs to compare
the physical properties and morphologies of these galaxies.

Several other studies of intermediate and high redshift galaxies
have found objects inconsistent with
rotation, similar to the DDGs.  Simard \& Pritchet (1998)
found 25\% of their galaxies had ``kinematically 
anomalous'' [O II] emission.  These objects had centrally
concentrated emission; we find below that DDGs are smaller in both
continuum and emission.  Erb \etal\ (2004) found that 
spectra of 13 elongated galaxies at $z\sim 2$ rarely show
rotation.  Bershady \etal\ (2005) obtained
HST/STIS spectra of six luminous compact blue galaxies and
found low $V/\sigma$; the STIS velocity
dispersion confirmed ground-based integrated linewidth.
From integral field spectroscopy, Flores \etal\ (2006) found that
$\sim 2/3$ of 35 galaxies at $0.4<z<0.75$ had either
mildly decentered or non-rotating kinematics; some of these
may be dispersion dominated.
Our TKRS survey establishes that the dispersion dominated
galaxies are relatively common and not restricted to
specially selected samples.

It is unclear what the local counterparts of dispersion
dominated galaxies could be.  Many local studies of galaxy
kinematics, in order to study the distance scale or rotation
curves and dark matter, have selected against galaxies with
disordered kinematics.  However, even very small galaxies
down to $M_B \sim -15-16$
are generally dominated by rotation (e.g. Swaters \etal\ 2002).
Tully-Fisher studies have 
considered the contribution of turbulent motions to the linewidth;
Tully \& Fouque (1985) claim that turbulent motions are only 
significant for very small rotation velocities $V \lesssim 25$ \kms,
but this is based on an idealized model.  Kannappan \& Barton (2004)
find a larger percentage of kinematically anomalous galaxies
in close pairs; these mostly show 
distorted or truncated rotation curves, but a few are unusual
enough that seeing-blur could make rotation gradients
undetectable.  It is also possible that evolution makes
bright dispersion dominated emission line galaxies rare at the
present day.

\subsection{Comparative Properties of Rotation and Dispersion 
Dominated Galaxies}

Rotation and dispersion dominated galaxies do exist over the 
full redshift range of our sample, as seen in Figure 
\ref{fig-zmagrotdispdom}.  There is some sign that at
$z \lesssim 0.4$, DDGs are only found at faint magnitudes.
This might reflect an evolution either in kinematics or
in magnitude and emission strength.  To compare the
properties of luminous RDGs and DDGs, and to avoid being
pulled by faint objects visible only at low redshift,
we show distributions of
galaxy properties for a sample restricted to $z>0.4$, and
with aligned slits and ellipticity $e>0.25$ as before.


\begin{figure}[ht]
\begin{center}
\includegraphics[width=3.5truein]{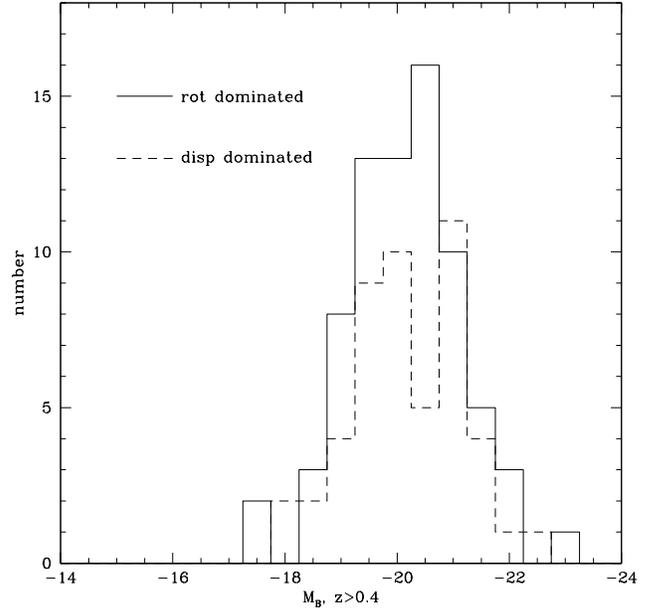}
\caption{
Magnitude distribution in restframe $M_B$ of rotation and dispersion 
dominated galaxies at $z>0.4$, with aligned slits and $e>0.25$.
Solid line: rotation dominated galaxies, $\vrot>\sigtwod$.
Dashed line: dispersion dominated galaxies, $\vrot<\sigtwod$.
At high redshift, RDGs and DDGs are equally bright.
}
\label{fig-rotdispmag}
\end{center}
\end{figure}

\begin{figure}[ht]
\begin{center}
\includegraphics[width=3.5truein]{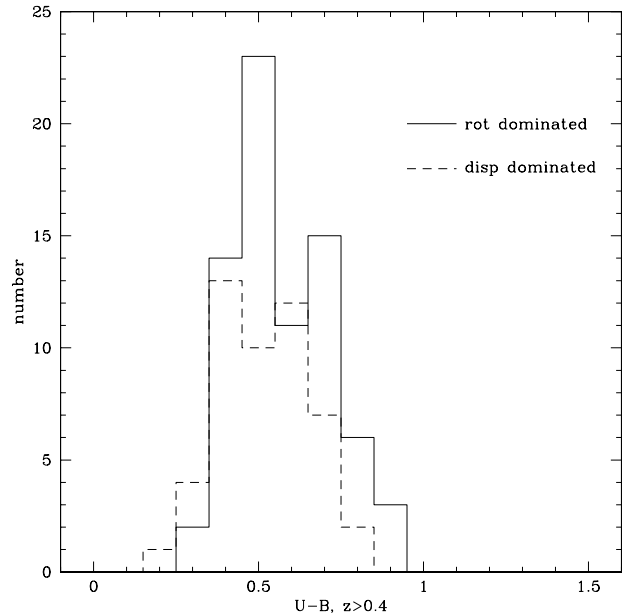}
\caption{
Color distribution in  restframe $U-B$ of rotation and dispersion 
dominated galaxies at $z>0.4$, with aligned slits and $e>0.25$.
Solid line: rotation dominated galaxies, $\vrot>\sigtwod$.
Dashed line: dispersion dominated galaxies, $\vrot<\sigtwod$.
RDGs are slightly redder than DDGs.
}
\label{fig-rotdispcolor}
\end{center}
\end{figure}


The distributions of $z>0.4$ RDG and DDG restframe magnitude and 
color $M_B$ and $U-B$  are shown in Figures \ref{fig-rotdispmag}
and \ref{fig-rotdispcolor}.  At $z>0.4$, the two types have
indistinguishable magnitude distributions.  The DDGs are 
bluer in the median than the RDGs, but only by 0.05 mag.
Comparing the RDG and DDG distributions with the Kolmogorov-Smirnov
statistic gives a 93\% probability that they are identical in
$M_B$ and a 15\% probability that they are identical in $U-B$.
The DDG color distribution covers almost the whole of the
range of the blue galaxy population, so it appears that
DDGs are not solely unusual blue objects, e.g. compact \hii\
regions or extreme starbursts.


\begin{figure}[ht]
\begin{center}
\includegraphics[width=3.5truein]{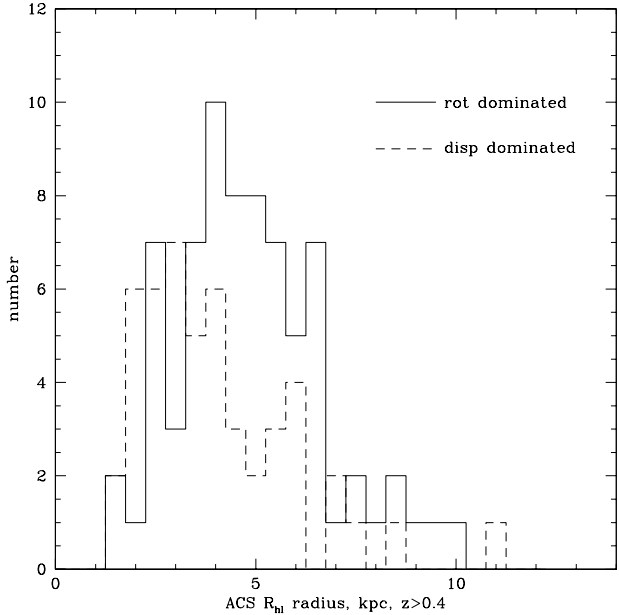}
\caption{
Radius distribution in half-light radius $R_{hl}$, kpc, from HST images,
for rotation and dispersion dominated galaxies at $z>0.4$,
with aligned slits and $e>0.25$.
Solid line: rotation dominated galaxies, $\vrot>\sigtwod$.
Dashed line: dispersion dominated galaxies, $\vrot<\sigtwod$.
RDGs are larger in the mean than DDGs, though the size distributions
overlap.
}
\label{fig-rotdispradius}
\end{center}
\end{figure}

\begin{figure}[ht]
\begin{center}
\includegraphics[width=3.5truein]{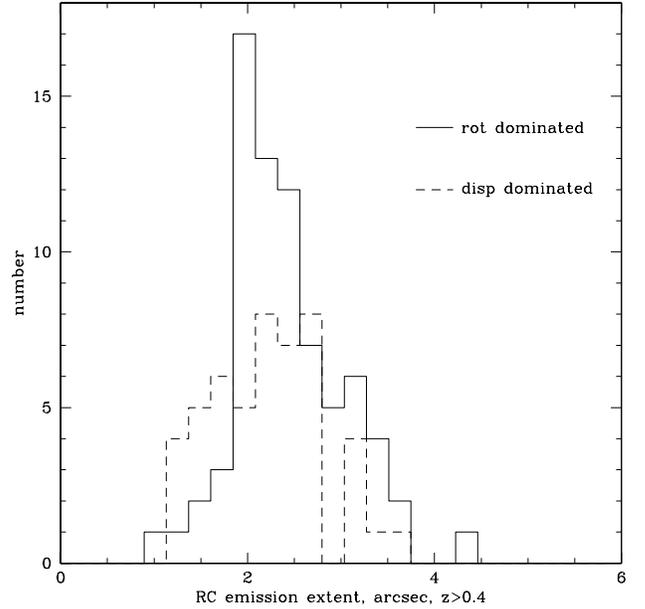}
\caption{
Emission extent: the total range (diameter), in arcsec, of data used in the 
rotation/dispersion fit in the Keck spectra,
for rotation and dispersion dominated galaxies at $z>0.4$,
with aligned slits and $e>0.25$.
Solid line: rotation dominated galaxies, $\vrot>\sigtwod$.
Dashed line: dispersion dominated galaxies, $\vrot<\sigtwod$.
RDGs are larger in the mean than DDGs, though less so than in 
half-light radius.  The ranges in emission extent are similar,
implying that DDGs are real and not just objects too small
to detect rotation in ground-based spectra.
}
\label{fig-rotdispemsize}
\end{center}
\end{figure}


The size distributions of $z>0.4$ RDGs and DDGs are shown in
continuum and in emission, in
Figures \ref{fig-rotdispradius} and \ref{fig-rotdispemsize}.
Figure \ref{fig-rotdispradius} plots the physical 
half-light radius $R_{hl}$ in kpc from the HST ACS $I$ images,
while Figure \ref{fig-rotdispemsize} plots the observed emission 
diameter in arcsec, $D_{em}$, using the range of data
in the Keck spectra deemed good by {\sc ROTCURVE} fitting.

Here there is a difference between DDGs and RDGs.  Dispersion
dominated galaxies are
smaller in both size measures.  The offset is more noticeable
in the physical radius $R_{hl}$.  Although the range of sizes
for DDGs is about as broad as for RDGs, the DDGs are
somewhat smaller in the average: the medians of $R_{hl}$
are 3.7 and 4.7 kpc for DDGs and RDGs.  The DDGs are similar in 
luminosity to RDGs but smaller in radius, so they are somewhat
more compact.

The emission diameter $D_{em}$ plotted in \ref{fig-rotdispemsize}
is the angular extent from the first to last row (inclusive)
in the Keck spectrum of the data that {\sc ROTCURVE} deemed good 
for fitting. There is a mild offset, with rotation dominated galaxies 
being larger in the average, but the RDGs and DDGs span the same
range of angular sizes.  The medians of $D_{em}$ are 2.26\arcsec\
and 2.38\arcsec\ for DDGs and RDGs.  The distribution of $D_{em}$
is limited at the small end because the rotation curve sample was 
selected by enforcing a lower limit in ground-based angular size
of FWHM = 0.94\arcsec.  
The K-S statistic gives a 2\% probability that the RDG and DDG
distributions are identical in $R_{hl}$ and a 12\% probability that
they are identical in $D_{em}$.  From the K-S tests, the only firm
difference between RDG and DDG properties is in the physical 
radius $R_{hl}$.
The two types of galaxy are not as different in $D_{em}$ as they are
in HST-measured physical radius; this could be caused by the selection
on ground-based size; by seeing blur; by differences in the falloff of
the light profiles; and/or by threshold effects, since $D_{em}$
measures where the emission strength drops below some value that
yields a reliable velocity.

The chief point of Figure \ref{fig-rotdispemsize} is that 
the DDGs and RDGs in our rotation curve sample span a similar
range of angular extent as measured on the 2-d spectra.  This 
implies that the dispersion dominated galaxies are genuinely 
non-rotating in their mean emission.  It argues against the
possibility that DDGs are simply an observational bias 
in small galaxies, caused by emission which does not
extend far enough to detect rotation, alluded to in
Section \ref{sec-rotdispdom}.  However, such a bias
might re-appear in samples without an angular size lower limit.

\subsection{Morphologies of Rotation and Dispersion 
Dominated Galaxies}

To explore further any systematic differences between
rotation and dispersion dominated galaxies, we made a
morphological classification of the sample of 116 RDGs
and 67 DDGs that have aligned slits and ellipticity $>0.25$.
We inspected ACS $I$-band images from the GOODS-N data.
The classifications are cataloged in Table \ref{table-rotcatalog}.

Because the physical resolution is limited and 
high redshift galaxies are known to exhibit a
greater degree of morphological peculiarity (e.g. 
Abraham \etal\ 1996), we defined a simplified system of 
morphological types.  We found no spheroidals in the RDG/DDG sample.
The classes are:
disks, Sb-Sd; irregulars, both low and high surface brightness;
edge-on disks; ``chain'' and ``hyphen'' galaxies; and mergers.
Chain galaxies are elongated objects with bright knots 
(e.g. Cowie, Hu, \& Songaila 1995).
Hyphen galaxies are thin, small but elongated, 
with little substructure.  We only defined 
galaxies as mergers if they clearly had multiple components,
tidal tails, or major disturbances associated with mergers.
Many high-$z$ galaxies are somewhat peculiar or asymmetric
and we did not count these as evidence of mergers.
The percentages of morphological types for RDGs and DDGs
are listed in Table \ref{table-morphtypes}, for both
the full sample and the sample restricted to $z>0.4$,
where RDGs and DDGs have similar luminosity.

\begin{deluxetable}{lrrrr}

\tablecaption{
Morphologies of rotation and dispersion dominated galaxies
\label{table-morphtypes}
}

\tablecolumns{6}
\tablewidth{0pt}

\tablehead{
Type  &  RDG     & RDG     & DDG     & DDG     \\
      &  all $z$ & $z>0.4$ & all $z$ & $z>0.4$
 }
 \startdata
spheroidal     &   0\%  &   0\%   &    0\%   &   0\% \\
disk           &  42\%  &  45\%   &   22\%   &  27\% \\
edge-on disk   &  16\%  &   8\%   &    2\%   &   0\% \\
irregular      &  31\%  &  32\%   &   43\%   &  41\% \\
chain/hyphen   &   7\%  &  10\%   &   25\%   &  22\% \\
merger         &   3\%  &   5\%   &    8\%   &  10\% \\
total          & 116    &  74     &   67     &  49   \\

 \enddata


\end{deluxetable}

The clearest results from Table \ref{table-morphtypes} are
that rotation dominated galaxies are generally disky, or
irregular, and very rarely chain or hyphen galaxies.  
Dispersion dominated galaxies sometimes appear
disky, but are more often irregular, and a substantial
number are chain or hyphen galaxies.  Hardly any DDGs
were edge-on disks, which is understandable since these
are most likely to have detectable rotation.  Mergers are 
not a large fraction of either RDGs or DDGs.

Restricting the sample to $z>0.4$ changes the ratios
little, other than to decrease the number of edge-on disk RDGs.
There is a definite offset in morphology between RDGs
and DDGs, but also substantial overlap.  These morphologies
are subjective; the link between kinematic and photometric 
morphologies will be addressed further with larger samples and 
objective measures in the DEEP2 survey.

A significant number of galaxies were classed as chain
or hyphen, types which are rare at low redshift.  Two-thirds 
of the chain/hyphen galaxies were DDGs; only one-third of
these elongated objects were dominated by rotation.  TKRS
5627 shown in Figures \ref{fig-exampleimages} and \ref{fig-examplerc}
is an example of a chain galaxy, elongated but nonrotating.  Similar
objects have been found at $z \sim 2$ by Erb \etal\ (2004): 
their slit spectra of 13 elongated galaxies showed dispersion,
but rarely rotation.   These results suggest that kinematic
surveys must use caution in interpreting ellipticity:
elongated objects may not be inclined disks and may not be
rotating.

The higher fraction of peculiar objects at high redshift suggests that
galaxies evolve in photometric morphology, with a greater fraction
of seemingly irregular galaxies at $z\sim 1$ (e.g. Abraham \etal\ 1996;
Lotz \etal\ 2006).
Possibly they also evolve in kinematic morphology.  Luminous
dispersion dominated galaxies appear at $z \sim 1$ but are uncommon
locally; even faint local blue galaxies are usually dominated by
rotation (e.g. Swaters \etal\ 2002).
Perhaps DDGs settle from a disordered or non-planar kinematic state 
into more orderly rotation; or their disordered kinematics are dominated
by bright star forming regions which then fade to allow a more
ordered background to be seen;  or they evolve in luminosity by 
fading more than the overall population; or they cease forming stars 
and both fade and drop out of emission-line kinematic samples,
leaving behind dispersion-supported faint red galaxies.

\section{Conclusions}
\label{sec-conclude}

We have measured line-of-sight kinematic linewidths \sigoned\ 
for $\sim 1000$ galaxies in the Team Keck survey of the GOODS-N field 
and spatially resolved line-of-sight 
rotation (\vrot) and dispersion (\sigtwod) profiles for 380
of these.  Most galaxies with linewidths are on the blue side of the
color bimodality.  Linewidths from integrated spectra are a measure of
internal kinematics that is relatively robust against observational
effects, based on simulations of
velocity fields and on comparisons of the linewidths with the
spatially resolved kinematics.  However, the unknown geometries of
high-redshift velocity fields, scatter in galaxy ellipticities,
and the lack of correction for projection into radial velocities
mean that there is a fair amount of scatter between an individual galaxy 
linewidth measurement and a true circular velocity or dynamical mass 
estimate.  For rotating 
galaxies, $\sigoned \sim 0.6 V_c$ in the mean, due to a combination 
of the inclination and the fact that $\sigoned$ is less than 
line-of-sight $\vrot$; this factor should be kept in mind when 
estimating dynamical masses.

The rotation and dispersion profiles of spatially resolved galaxies
show that not all galaxies
exhibit a conventional rotation curve.  Galaxies can be roughly
divided into rotation and dispersion dominated, with $\vrot>\sigtwod$
or vice versa.  Dispersion dominated galaxies are blue, mostly
irregular, and are not ellipticals; they probably have an effective
dispersion, from disordered kinematics which are integrated over by the 
seeing.  Dispersion dominated galaxies exist 
at all redshifts in our sample, but low-$z$ DDGs are quite faint.

When line-of-sight rotation and dispersion are combined to make an
estimate of the overall dispersion of the potential, as 
in $\whalf^2 = 0.5\vrot^2 + \sigtwod^2$, the result correlates
well with the integrated linewidth \sigoned, demonstrating
both that \sigoned\ is a robust velocity indicator and that
it is possible to construct scaling relations with velocity
for a population of diverse kinematic properties.  In Paper II
we use the linewidths to measure evolution in the Tully-Fisher
relation.

At $z \sim 1$, rotation and dispersion dominated galaxies have similar
magnitudes and colors, but somewhat different size and morphology
distributions.  Rotation dominated galaxies are mostly disk and
irregular types; dispersion dominated galaxies are in the mean smaller
in physical half-light radius, and $\sim 60\%$ of DDGs are irregular
and chain/hyphen types.  Only a small fraction of either RDGs or DDGs
are obvious mergers.  About two-thirds of chain and hyphen galaxies
are dispersion dominated: elongated high-redshift objects cannot be
assumed to be inclined rotating disks.  It is not clear what the local
counterparts of dispersion dominated galaxies might be; integral field
spectroscopy, especially with adaptive optics, may shed some light on
their nature at high redshift.

There is some evidence that dispersion dominated galaxies are rare at
bright magnitudes at low redshift, just as previous studies have shown
that irregular photometric morphologies in bright galaxies are more
common at high redshift than locally (e.g. Abraham \etal\ 1996;
Lotz \etal\ 2006).  It is possible that galaxies
evolve in kinematic morphology, settling into ordered rotation, or
that luminosity evolution causes them to appear more ordered, or to
drop out of bright emission-line samples.  These and other
possibilities are avenues for further study with larger samples and
quantitative morphological measurements.

\acknowledgments
We dedicate this paper to the memory of Bob Schommer.
We thank the TKRS, GOODS, and Hawaii groups for making
their catalogs and data publicly available.  
We are grateful to Ted Williams and Bob Schommer for observing 
NGC 7171 with the Rutgers Fabry-Perot.
BJW has been supported by grant NSF AST-0242860 to S. Veilleux.  
The TKRS was supported by NSF grant AST-0331730 and this 
project has been supported by NSF grant AST-0071198 to UCSC.  
AJM acknowledges support from NSF grant AST-0302153 through the 
NSF Astronomy and Astrophysics Postdoctoral Fellows Program.
The authors wish to recognize and acknowledge the 
cultural role and reverence that the summit of Mauna Kea
has always had within the indigenous Hawaiian community.
We are most fortunate to have the opportunity to conduct
observations from this mountain.



\begin{references}

\bigskip

{\small

\reference{} Abraham, R.G., van den Bergh, S., Glazebrook, K., 
Ellis, R.S., Santiago, B.X., Surma, P., \& Griffiths, R.E.\ 1996, \apjs, 107, 1


\reference{} Bamford, S.P., Milvang-Jensen, B., Arag{\'o}n-Salamanca, A., 
\& Simard, L.\ 2005, \mnras, 361, 109

\reference{} Bamford, S.P., Arag{\'o}n-Salamanca, A., \&
Milvang-Jensen, B.  2006, \mnras, 366, 308

\reference{} Barton, E.J., Geller, M.J., Bromley, B.C., 
van Zee, L., \& Kenyon, S.J.\ 2001, \aj, 121, 625 

\reference{} Barton, E.J., \& van Zee, L.\ 2001, \apjl, 550, L35 



\reference{} Bell, E. F., Wolf, C., Meisenheimer, K., Rix, H.-W., 
Borch, A., Dye, S., Kleinheinrich, M., \& McIntosh, D. H. 2004, ApJ, 608, 752

\reference{} Bershady, M.A., Vils, M., Hoyos, C., Guzman, R., \& Koo, D.C.
2005, in ASSL Vol. 329: 
Starbursts: From 30 Doradus to Lyman Break Galaxies, 177

\reference{} Bertin, E., \& Arnouts, S.\ 1996, \aaps, 117, 393

\reference{} Binney, J., \& Tremaine, S.  1987, Galactic Dynamics, 
(Princeton: Princeton U.P.)


\reference{} B{\" o}hm, A., et al.\ 2004, \aap, 420, 97 



\reference{} Capak, P., et al.\ 2004, \aj, 127, 180 

\reference{} Conselice, C.J., Bundy, K., Ellis, R.S., Brichmann, J., 
Vogt, N.P., \& Phillips, A.C.\ 2005, \apj, 628, 160

\reference{} Cowie, L.L., Hu, E.M., \& Songaila, A. 1995, \aj, 110, 1576



\reference{} Davis, M. \etal\ 2003, Proc. SPIE, 4834, 161

\reference{} Erb, D.K., Steidel, C.C., 
Shapley, A.E., Pettini, M., \& Adelberger, K.L.\ 2004, \apj, 612, 122 


\reference{} Fioc, M., \& Rocca-Volmerange, B.\ 1997, \aap, 326, 950 

\reference{} Flores, H., Hammer, F., Puech, M., Amram, P. \& Balkowski, C.
2006, \aap, in press, astro-ph/0603563



\reference{} Giavalisco, M., et al.\ 2004, \apjl, 600, L93 

 
\reference{} Giovanelli, R., Haynes, M.P., Salzer, J.J., Wegner, G., 
da Costa, L.N., \& Freudling, W. 1995, \aj, 110, 1059

\reference{} Kannappan, S.J., \& Barton, E.J.\ 2004, \aj, 127, 2694 


\reference{} Kinney, A. L., Calzetti, D., Bohlin, R. C., McQuade, K., 
Storchi-Bergmann, T., \& Schmitt, H. R. 1996, \apj, 467, 38

\reference{} Kobulnicky, H.A., \& Gebhardt, K.\ 2000, \aj, 119, 1608 

\reference{} Koo, D.C., Guzman, R., Faber, S.M., Illingworth, G.D., 
Bershady, M.A., Kron, R.G., \& Takamiya, M.\ 1995, \apjl, 440, L49


\reference{} Lehnert, M.D., \& Heckman, T.M.\ 1996, \apj, 472, 546 



\reference{} Lotz, J.M. et al.\ 2006, \apj, submitted, astro-ph/0602088

\reference{} Mall{\' e}n-Ornelas, G., Lilly, S.J., Crampton, D., 
\& Schade, D.\ 1999, \apjl, 518, L83 


\reference{} Melbourne, J., et al.\ 2006, \apj, submitted, astro-ph/

\reference{} Metevier, A.J., Koo, D.C, Simard, L., \& Phillips, A.C.  
2006, \apj, 643, 764 

\reference{} Milvang-Jensen, B., Arag{\' o}n-Salamanca, A., Hau, G.K.T., 
J{\o}rgensen, I., \& Hjorth, J.\ 2003, \mnras, 339, L1 


\reference{} Nakamura, O., Arag{\' o}n-Salamanca, A., Milvang-Jensen, B.,
Arimoto, N., Ikuta, C., \& Bamford, S.P. 2006, \mnras, 366, 144


\reference{} Palunas, P.  1996, PhD thesis, Rutgers University

\reference{} Palunas, P., \& Williams, T.B.\ 2000, \aj, 120, 2884 

\reference{} Pisano, D.J., Kobulnicky, H.A., Guzm{\' a}n, R., 
Gallego, J., \& Bershady, M.A.\ 2001, \aj, 122, 1194 


\reference{} Press, W.H., Flannery, B.P., Teukolsky, S.A. \& Vetterling,
W.T. 1992, Numerical Recipes,  (Cambridge: Cambridge U.P.)

\reference{} Rix, H., Guhathakurta, P., 
Colless, M., \& Ing, K.\ 1997, \mnras, 285, 779 

\reference{} Sakai, S. et al.\ 2000, \apj, 529, 698




\reference{} Simard, L., \& Pritchet, C.J.\ 1998, \apj, 505, 96 



\reference{} Strateva, I., et al. 2001, AJ, 122, 1861

\reference{} Swaters, R.A., van Albada, T.S., van der Hulst, J.M., \& 
Sancisi, R.\ 2002, \aap, 390, 829



\reference{} Tully, R.B. \& Fouque, P.  1985, \apjs, 58, 67



\reference{} Tully, R.B., Pierce, M.J., Huang, J., Saunders, W., 
Verheijen, M.A.W., \& Witchalls, P.L.\ 1998, \aj, 115, 2264

\reference{} Vogt, N.P., et al.\ 1996, \apj, 465, L15

\reference{} Vogt, N.P., et al.\ 1997, \apj, 479, L121

\reference{} Vogt, N.P.\ 2000, ASP Conf.~Ser.~197: Dynamics of 
Galaxies: from the Early Universe to the Present, 197, 435


\reference{} Weiner, B.J., Williams, 
T.B., van Gorkom, J.H., \& Sellwood, J.A.\ 2001, \apj, 546, 916

\reference{} Weiner, B.J., et al.\ 2005, \apj, 620, 595

\reference{} Weiner, B.J., et al.\ 2006, \apj, (Paper II)



\reference{} Willmer, C.N.A., et al.\ 2006, \apj, in press

\reference{} Wirth, G.D., et al.\ 2004, \aj, 127, 3121


\reference{} Ziegler, B.L., et al.\ 2002, \apjl, 564, L69 

}

\end{references}
\end{document}